\documentclass[aps,pra,
showpacs,
showkeys,
numerical,
amsmath,amssymb,
reprint
]{revtex4-1}

\usepackage{amsmath}
\usepackage{xspace}
\usepackage{relsize}
\usepackage{tikz}
\usepackage[utf8]{inputenc}
\usepackage{bm} 
\usepackage{amssymb}
\usepackage{ctable}
\usepackage[version=3]{mhchem}
\usepackage{xr}
\newcommand{\python}  {\texttt{python}\xspace}
\newcommand{\julia}   {\texttt{julia}\xspace}

\newcommand{\molsturm}{\texttt{molsturm}\xspace}
\newcommand{\sturmint}{\texttt{sturmint}\xspace}
\newcommand{\ie}{\mbox{i.e.}\xspace}
\newcommand{\eg}{\mbox{e.g.}\xspace}

\renewcommand*{\vec}[1]{\ensuremath{\underline{\boldsymbol{#1}}}}
\newcommand*{\uvec}[1]{\ensuremath{\hat{\underline{\boldsymbol{#1}}}}}
\newcommand{\bigO}{\ensuremath{\mathcal{O}}}
\newcommand{\R}{\mathbb{R}}
\DeclareMathOperator{\spacespan}{span}

\newcommand{\SCF}{SCF\xspace}
\newcommand{\HF}{HF\xspace}
\newcommand{\UHF}{UHF\xspace}
\newcommand{\RHF}{RHF\xspace}

\newcommand{\DFT}{DFT\xspace}
\newcommand{\CS}{CS\xspace}
\newcommand{\cGTO}{cGTO\xspace}
\newcommand{\ETO}{ETO\xspace}
\newcommand{\STO}{STO\xspace}
\newcommand{\mat}[1]{\mathbf{#1}}
\newcommand{\RMS}{RMS\xspace}
\newcommand{\CBS}{CBS\xspace}
\newcommand{\kexp}{k}
\newcommand{\nmax}{\ensuremath n_\text{max}}
\newcommand{\lmax}{\ensuremath l_\text{max}}
\newcommand{\mmax}{\ensuremath m_\text{max}}
\newcommand{\Nbas}{N_\text{bas}}

\newcommand{\kopt}{\ensuremath k_\text{opt}}
\newcommand{\RMSOl}{\text{RMSO}_l}
\newcommand{\Itpf}{\mathcal{I}_F}
\newcommand{\Ibas}{\mathcal{I}_\text{bas}}

\newcommand{\laplaceRadial}{\frac{\partial}{\partial r} \left( r^2 \frac{\partial}{\partial r} \right)}

\begin{document}
\title{
	Quantum chemistry with Coulomb Sturmians:
	Construction and convergence of
	Coulomb Sturmian basis sets at
	Hartree-Fock level
}

\author{Michael F. Herbst}
\email{michael.herbst@iwr.uni-heidelberg.de}
\affiliation{
	Interdisciplinary Center for Scientific Computing, Heidelberg University, Im Neuenheimer Feld 205, 69120 Heidelberg, Germany
}

\author{James Emil Avery}
\email{avery@nbi.ku.dk}
\affiliation{
	Niels Bohr Institute, University of Copenhagen, Blegdamsvej 17, 2100 København, Denmark
}

\author{Andreas Dreuw}
\email{dreuw@uni-heidelberg.de}
\affiliation{
	Interdisciplinary Center for Scientific Computing, Heidelberg University, Im Neuenheimer Feld 205, 69120 Heidelberg, Germany
}

\begin{abstract}
	The first discussion
	of basis sets consisting of exponentially
	decaying Coulomb Sturmian functions
	for modelling electronic structures is presented.
	The proposed basis set construction
	selects Coulomb Sturmian functions using separate upper limits to
	their principle, angular momentum and magnetic quantum numbers.
	Their common Coulomb Sturmian exponent is taken as a fourth parameter.
	The convergence properties of such basis sets are investigated
	taking the second and third row atoms at Hartree-Fock level as examples.
	Thereby important relations between the values of the basis set parameters
	and the physical properties of the electronic structure are recognised.
	Furthermore, a connection between the optimal,
	\ie minimum-energy, Coulomb Sturmian exponent
	and the average Slater exponents values
	obtained by Clementi and Raimondi%
	~(E.~Clementi and D.~L.~Raimondi, J.~Chem.~Phys. 38, 2686 (1963))
	is made.
	These features of Coulomb Sturmian basis sets
	emphasise their ability to correctly reproduce
	the physical features of Hartree-Fock wave functions.
	As an outlook the application of Coulomb Sturmian discretisations
	for molecular calculations and Post-HF methods is briefly discussed.
\end{abstract}

\pacs{31.15.ac,31.15.ae,31.15.xj,31.15.xr}
\keywords{Coulomb Sturmians, basis set construction, Hartree-Fock, convergence properties}

\maketitle

\section{Introduction}
\label{sec:Introduction}
The standard approach for approximating solutions
to the electronic Schrödinger equation is to
employ a limited set of single-particle basis functions
to build a discretisation basis.
An early approach pursued by Slater~\cite{Slater1930}
was to employ exponential-type orbitals~(\ETO)
with a radial part of the form $\exp\left( -\zeta r\right)$ times a polynomial.
In his construction the exponent $\zeta$ was estimated from empirical rules,
but later refined exponents based on Hartree-Fock calculations
became available~\cite{Clementi1963}.
Whilst \ETO could thus be readily used for modelling atoms,
difficulties related to the evaluation of multi-centred two-electron
repulsion integrals~(ERI)
directed attention to other types of basis functions
for molecular calculations.
An outcome of this development are
contracted Gaussian-type orbitals~(\cGTO)%
~\cite{Huzinaga1965,Hehre1969},
for which the evaluation of ERI is much simpler due to the Gaussian product theorem.
Over the years many kinds of \cGTO basis sets
have been developed,~\cite{Jensen2013,Hill2013}
such that now most aspects of electronic structure
can be modelled reliably using \cGTO functions.

Compared to an \ETO basis a missing aspect of \cGTO basis sets is, however,
that these functions are not able to correctly reproduce
the functional form~\cite{Kato1951,Kato1957} of the wave function
at large distances or close to the nucleus.
For properties such as
nuclear-magnetic resonance (NMR) shielding tensors~\cite{Guell2008,Hoggan2009}
or Rydberg-like or auto-ionising states%
~\cite{Feshbach1958,Feshbach1962,Riss1993,Santra2002},
where either the nuclear cusp or the asymptotic tail
are important~\cite{Guell2008,Hoggan2009},
\ETO basis sets remain attractive.
Additionally,
the computational resources available for quantum-chemical calculations
have changed since the 1970's,
such that it may now be favourable to invest extra computation
per integral in order to have fewer, more accurate basis functions.

Following the pioneering efforts by
Harris, Michels, Steinborn, Weniger, Weatherford, Jones, and others%
~\cite{HarrisMichels1967,Steinborn1983,Weniger1983,Weatherford1982},
in making ETOs more efficient,
recently Coulomb Sturmians%
~\cite{Shull1959,Rotenberg1962,Rotenberg1970,Aquilanti2003,Coletti2013,Calderini2012}~(CS)
have emerged as a particularly promising \ETO basis.
Firstly, the momentum-space representation of these functions
is equivalent to the hyperspherical harmonics,
which allows multi-centre electron repulsion integrals
to be treated rather efficiently%
~\cite{Avery2012,Avery2013,Avery2015,Avery2017}.
This opens way for treating molecular problems based on these \ETO in the future.
Secondly, understanding \CS basis sets provides a foundation
towards the investigation of other \ETO basis function types,
since the Coulomb Sturmian construction can be easily generalised
preserving many useful properties of the \CS functions~\cite{Avery2006}.
For example, one may build $N$-particle basis functions
that include geometric properties
of the physical system under consideration
at the level of the basis%
~\cite{Aquilanti1996,Aquilanti1997,Aquilanti1998,Avery2003,Avery2004,Avery2006,Avery2009,Calderini2013,Abdouraman2016}.
Similarly, $d$-dimensional hyperspherical harmonic basis sets can model collective
motions of particles, for example for treating strongly interacting few-body systems 
or reactive scattering%
~\cite{Avery1989-hyperbook,Aquilanti1992,Aquilanti2004,Avery2018-hyperbook,Das2016}.
With respect to scattering problems
employing Coulomb Sturmians and generalised Sturmians
has become a well-established technique~%
\cite{Randazzo2010,Mitnik2011,Ambrosio2015,GranadosCastro2015,Randazzo2015,Granados2016}
and construction schemes for optimal
Sturmian bases have been suggested~\cite{Randazzo2010}.

In a recent publication we presented the \molsturm framework~\cite{molsturmDesign}
in which atomic electronic-structure calculations
based on a Coulomb Sturmian discretisation can be performed
both at Hartree-Fock~(\HF) and Post-\HF level.
Unlike the application of Sturmians to scattering
and unlike conventional \cGTO discretisations,
construction schemes for reliable and efficient \CS basis sets
are not yet available to the best of our knowledge.
The aim of this paper is to provide a first step towards closing
this gap, allowing to readily conduct \CS-based electronic-structure calculations
in the future.
In particular, this work is concerned with the construction
of \CS basis sets for atomic systems at Hartree-Fock level,
which represents first elementary step for the construction of molecular basis sets.
Appropriate modifications for capturing electronic correlation
and to compute excited states will be briefly hinted at in the outlook.
However, more detailed discussion on this matter is planned for
a follow-up publication.

\subsection{Coulomb Sturmian basis functions}
\label{sec:CSIntro}
Coulomb Sturmians are the analytic solutions to the
single-particle equation~\cite{Avery2006}
\begin{equation}
	\left( - \frac12 \Delta - \beta_n \frac{Z}{r} - E \right) \varphi^\text{CS}_\mu(\vec{r}) = 0.
	\label{eqn:CS}
\end{equation}
This equation can be considered as a modification of the hydrogen-like Schrödinger
equation, where the Coulomb attraction between electron and nucleus
is scaled by a factor
\begin{equation}
	\beta_n = \frac{\kexp n}{Z}.
	\label{eqn:CSbeta}
\end{equation}
Separation of \eqref{eqn:CS} into radial and angular variables,
yields the \textit{Coulomb Sturmian radial equation}
\begin{equation}
	\left( - \frac{1}{2r^2} \laplaceRadial + \frac{l (l+1)}{2 r^2}
	- \frac{n\kexp}{r} - E \right) R_{nl}(r) = 0.
	\label{eqn:CSradial}
\end{equation}
Equation \eqref{eqn:CSradial} defines the \CS radial part $R_{nl}$,
which is identical to the familiar hydrogen-like orbitals,
just with all occurrences of the factors $Z/r$ replaced by
the Coulomb Sturmian exponent $k$.
The full functional form of the \CS reads
\begin{equation}
	\begin{aligned}
	\varphi_{nlm}(\vec{r}) &= R_{nl}(r) Y_l^m(\uvec{r}), \\
	R_{nl}(r) &= k^{3/2} N_{nl} (2k r)^l e^{-k r} L^{2l+1}_{n-l-1}(2k r),
	\end{aligned}
	\label{eqn:CSdef}
\end{equation}
where $Y_l^m$ is a spherical harmonic,
$L^{2l+1}_{n-l-1}$ an associated Laguerre polynomial and
\begin{equation}
N_{nl} = \frac{2}{(2l+1)!} \sqrt{ \frac{(l+n)!}{n (n-l-1)!}}
\end{equation}
the normalisation constant.
Next to other prominent atom-centred basis functions
such as Slater-type orbitals~(\STO) or Gaussian-type orbitals,
these functions share the factorisation into
a radial part, $R_{nl}$, and a spherical harmonic $Y_l^m$.
In contrast to \STO, however,
all \CS functions in a \CS basis set share the exponent $k$,
which furthermore is related to the
energy eigenvalue of Eq.~\eqref{eqn:CS}
\begin{equation}
	E = -\frac{\kexp^2}{2}.
	\label{eqn:CSenergy}
\end{equation}

Similar to the Schrödinger equation for hydrogen-like atoms,
Eq.~\eqref{eqn:CS} can only be solved for some
quantum number triples $(n, l, m)$, namely those from the set
\begin{equation}
	\begin{aligned}
	\Itpf \equiv \Big\{ (n, l, m) \, \Big| \, &n,l,m \in \mathbb{Z} \quad \text{with}\\
	&n > 0, \  0 \leq l < n,\  -l \leq m \leq l\Big\}.
	\end{aligned}
	\label{eqn:AllCSTriples}
\end{equation}
Furthermore one follows the convention to call
$n$, $l$ and $m$ the principle, angular momentum and magnetic
quantum numbers and uses both the spectroscopic terminology
$1s$, $2s$, $2p$, $\ldots$
as well as the corresponding quantum number triple
to refer to a particular \CS function.

The Coulomb Sturmian radial Eq.~\eqref{eqn:CSenergy}
is of Sturm-Liouville form~\cite{Rotenberg1970,Avery2006},
equipping \CS basis functions with some noteworthy properties.
Firstly, building on the arguments of
Klahn and Bingel~\cite{Klahn1977,Klahn1977a} one can show~\cite{Avery2008}
the countably infinite set of all Coulomb Sturmians
$\{\varphi^\text{CS}_\mu\}_{\mu\in\Itpf}$
to be a complete basis for the Sobelev space $H^1(\R^3)$ \emph{independent}
of the value of the exponent $\kexp$.
This is both the relevant Hilbert space for solving
the one-particle Schrödinger equation~\cite{Teschl2014}
as well as the Hartree-Fock problem for many-body systems.
As a consequence the numerical challenges
associated with treating
high-energy Rydberg-like, dipole-bound or ionising states
are most likely less pronounced
with a \CS-based approach.

Furthermore the Sturm-Liouville form of the radial equation
\eqref{eqn:CSradial} implies that the radial parts $R_{nl}$
form a complete basis for each value of $l$.
We will employ this to design \CS basis sets,
which subsequently converge the radial part,
but do not tighten the angular discretisation beyond an initial level.
Such basis set progressions can be used to understand
the maximal angular momentum quantum number
required in a \CS basis set
for describing the wave function
at a particular level of theory.

\subsection{The importance of selecting angular momentum quantum numbers
in quantum-chemical basis sets}
\label{sec:ImportanceAM}
Understanding which angular momentum quantum numbers
are required in a basis
is not a question limited to Coulomb Sturmians.
Much rather this aspect is of general concern
when constructing atom-centred basis sets for quantum-chemical modelling.
In the familiar context of \cGTO basis functions,
for example,
all basis sets used for practical calculations
include at least polarisation functions,
\ie functions whose angular momentum quantum number
exceeds the minimal basis set value.
This allows both to capture the density reorganisation
going from atoms to molecular structures
as well as the leading order effects of electronic correlation%
~\cite{Francl1982,Krishnan1980,Hariharan1973,Kong2012,Jensen2013}.
Similarly the systematic construction of \cGTO basis sets
with steady and reliable convergence behaviour
is closely related to selecting the amount of angular momentum
to be included%
~\cite{Dunning1989,Jensen2012}.

Additionally, investigating the required angular momentum quantum numbers
can become a diagnostic tool.
An example is the unphysical breaking of spherical symmetry
in the unrestricted \HF~(\UHF) modelling of atoms.
When considering the results
of an \UHF calculation of carbon and fluorine, \citet{Cook1981} noticed the
$s$-type and $p$-type \HF orbitals
for both these systems not to be linear combinations
of \cGTO basis functions with $l=0$ and $l=1$,
but to involve functions with higher angular momentum quantum numbers as well.
This was later found to be a general issue of
\UHF~\cite{Fukutome1981,Cook1984,McWeeny1985}.
Reference \onlinecite{Fukutome1981} provides a detailed analysis of the
underlying mechanisms including a discussion of the effect of
symmetry breaking and \HF instabilities in \UHF
and other \HF variants.

Understanding which angular momentum quantum numbers
a basis set needs to provide
is thus of general importance
to treat problems in molecular quantum chemistry
and to understand their properties.
With respect to Coulomb Sturmian discretisations
this work will discuss the
root-mean-square occupied coefficient value
per angular momentum%
~($\RMSOl$) to demonstrate the capabilities of such discretisations.
As will be discussed
this quantity allows to directly observe the oddities
with respect to the \UHF-induced breaking of spherical symmetry.
Furthermore in general the angular momentum requirements
of \HF wave functions can be directly probed in this way.
Similar to \cGTO discretisations our obtained results represent the
first step towards constructing more general CS basis sets
for correlated methods or molecular calculations.

\subsection{Paper Outline}
The remainder of the paper is structured as follows:
Section \ref{sec:Theory} introduces the theoretical background
and describes the computational methodologies.
The obtained \HF convergence results
are discussed in section \ref{sec:Results}.
Section \ref{sec:Outlook} provides an outlook
towards Post-\HF methods and other directions of future work.

\section{Theory and methodology}
\label{sec:Theory}

\subsection{Parameters for denoting Coulomb Sturmian basis sets}
\label{sec:BasisCS}
As outlined in section \ref{sec:CSIntro}, all Coulomb Sturmian functions
share a common exponent $\kexp$, but differ in the
quantum numbers $n$, $l$ and $m$,
which are taken from the set $\Itpf$ (see Eq.~\eqref{eqn:AllCSTriples}).
A Coulomb Sturmian basis set is therefore uniquely defined
by denoting the selection of triples $(n, l, m) \in \Itpf$ employed
as well as the value for the exponent $\kexp$.

Theoretically any selection of triples $(n, l, m) \in \Itpf$
can be used to form a \CS basis.
From the similarity of the \CS functions to the hydrogen-like orbital functions
one would, however, one can expect Coulomb Sturmians with smaller values of
the principle quantum number $n$ to be most important.
Both chemical intuition as well as the typical construction schemes
employed for \cGTO basis sets~\cite{Jensen2013} suggest to additionally limit
the angular momentum quantum number $l$ from above as well.
Guided by these ideas we focus in our investigation on \CS basis sets of the form
\begin{equation}
	\begin{aligned}
	\Ibas \equiv
\big\{ \varphi_{nlm} \, \Big| \,  (n, l, m) \in \Itpf, \   &n \le \nmax,
	\  l \le \lmax, \\
	&-\mmax \le m \le \mmax \big\},
	\end{aligned}
	\label{eqn:CSBasisSet}
\end{equation}
\ie where all three quantum numbers are bound from above.
For ease of notation, we will refer to \CS basis sets like
Eq.~\eqref{eqn:CSBasisSet} by the triple $(\nmax, \lmax, \mmax)$.
For example a $(3,2,2)$ \CS basis denotes the set with
$\nmax = 3$, $\lmax = 2$ and $\mmax = 2$.
Typically we will not place explicit bounds on all three quantum numbers.
For example $m$ is usually not restricted beyond the
limit $|m| < l$ intrinsic to the \CS equation~\eqref{eqn:CS}.
We will refer to such a basis as being only restricted by $\nmax$ and $\lmax$.
Similarly a basis only restricted by $\nmax$
has no tighter bound on $l$ apart from the condition $l < n$ already encoded in $\Itpf$.

Tuning the maximal quantum numbers $\nmax$, $\lmax$ and $\mmax$
naturally influences the subset of radial functions $R_{nl}$
and spherical harmonics $Y_l^m$,
which is available in the discretisation basis.
Since the Coulomb Sturmian radial equation \eqref{eqn:CSradial}
is of Sturm-Liouville form, the eigenfunctions of Eq. \eqref{eqn:CSradial},
\ie the set of all radial functions $\{R_{nl'}\}_{n>0}$ with $l'$ fixed,
is a complete basis for a weighted $L^2$ space~\cite{Avery2008}.
This allows to express each function $R_{nl}$ with arbitrary $l$
as a linear combination of functions $\{R_{n',l'}\}_{n'>0}$
of a different $l'$.
By considering the polynomial spaces spanned by
the \CS radial functions one can show
that for given $n$ and $l$ a set consisting only of the radial parts with
$l'=0$ and $n' \leq n$ is sufficient to form $R_{nl}$.
As a result
\begin{equation}
	\begin{aligned}
	&\forall \nmax > 0, 0 \leq l < \nmax: \\
		&\hspace{20pt}
	\spacespan \{R_{n'l'}\}_{n' \leq \nmax, l' \leq l}
		= \spacespan \{R_{n',0}\}_{n'\leq \nmax}.
	\end{aligned}
	\label{eqn:RadialSpans}
\end{equation}
Convergence in the radial discretisation in a \CS basis
can thus be completely controlled by tuning the bound $\nmax$.
Conversely $\lmax$ and $\mmax$ only effect convergence with respect to the angular part
in agreement with the physical interpretation given to the quantum numbers $l$ and $m$.
Notice, that these arguments are independent of the value of $\kexp$,
and as such apply for any value of the \CS exponent.

Provided that a value for $\lmax$ has been found,
which provides a good enough angular discretisation,
additional convergence in the radial coordinates can
therefore be achieved by only increasing the bound $\nmax$ of the \CS basis.
The implied strategy,
namely to restrict both $\nmax$ and $\lmax$
possesses the additional advantage
that the scaling of the basis size with respect to $\nmax$ is reduced
compared to only restricting $\nmax$.
Explicitly, in the latter case the resulting basis consists of
\begin{equation}
	\begin{aligned}
	\Nbas(\nmax) &= \sum_{n=1}^{\nmax} \sum_{l=0}^{n-1} \sum_{m=-l}^l 1
		= \sum_{n=1}^{\nmax} \sum_{l=0}^{n-1} 2l+1 \\
		&= \sum_{n=1}^{\nmax} n^2
		= \frac{(2\nmax+1)(\nmax+1)\nmax}{6} \\
		&\in \bigO(\nmax^3),
	\end{aligned}
	\label{eqn:NbasFullShellBasis}
\end{equation}
functions, \ie scales cubically with $\nmax$.
In comparison an additional restriction by $\lmax$ leads to
\begin{equation}
	\begin{aligned}
	\Nbas(\nmax) &= \sum_{n=1}^{\nmax} \ \sum_{l=0}^{\min(\lmax, n-1)} 2l + 1 \\
	&= \sum_{n=1}^{\nmax} \Big( \min(\lmax+1, n) \Big)^2 \\
	&\le \sum_{n=1}^{\nmax} (\lmax+1)^2 = (\nmax-1) (\lmax+1)^2 \\
	&\in \bigO(\nmax \lmax^2),
	\end{aligned}
	\label{eqn:NbasAmLimitedBasis}
\end{equation}
\ie linear scaling in $\nmax$.
Since the prefactor depends on $\lmax^2$, however,
a small value for $\lmax$ is desirable.

\subsection{Hartree-Fock variants and fractional occupation scheme}
\label{sec:HFfractional}
All variants of Hartree-Fock~(\HF)~\cite{Roothaan1951,Pople1954,Szabo1996}
can be viewed as a minimisation procedure of an appropriate energy functional
with respect to the occupied \HF or \DFT orbitals%
~\cite{Cances2000,Cances2000a,Cances2000b}.
Employing a finite-sized basis set for the
discretisation and separating the
resulting Euler-Lagrange equations
into $\alpha$ and $\beta$ spin components,
yields the following system of coupled non-linear eigenvalue problems:
\begin{equation}
	\begin{aligned}
	\mat{F}^\sigma\left( \mat{D}^\alpha, \mat{D}^\beta \right) \mat{C}^\sigma
	&= \mat{S} \mat{C}^\sigma \mat{E}^\sigma, \\
	\mat{C}^{\sigma \dagger} \mat{S} \mat{C}^\sigma &= \mat{I},
	\end{aligned}
	\label{eqn:SCFeulerLagrange}
\end{equation}
where $\sigma \in \{\alpha, \beta\}$ indicates the spin component,
$\mat{C}^\sigma$ are the matrices of orbital coefficients,
$\mat{E}^\sigma$ the diagonal matrices of orbital energies,
$\mat{S}$ is the overlap matrix and $\mat{I}$ the identity matrix.
The non-linearity of \eqref{eqn:SCFeulerLagrange} originates from
the Fock matrix, since $\mat{F}\left( \mat{D}^\alpha, \mat{D}^\beta \right)$
depends on both densities $\mat{D}^\sigma$.
These in turn are related to the coefficients $\mat{C}^\sigma$ via
\begin{equation}
	\mat{D}^\sigma = \mat{C}^\sigma \mat{f}^\sigma \mat{C}^{\sigma \dagger},
	\label{eqn:DensityMatrix}
\end{equation}
where $\mat{f}^\sigma$ is the diagonal matrix of occupation numbers.
Equations \eqref{eqn:SCFeulerLagrange} may be solved iteratively
employing the \textit{self-consistent field procedure}~(\SCF)~\cite{Roothaan1951}.
For restricted closed-shell \HF~(\RHF)~\cite{Roothaan1951}
one takes $\mat{F}^\alpha = \mat{F}^\beta$,
which implies $\mat{D}^\alpha = \mat{D}^\beta$.
Thus Eq.~\eqref{eqn:SCFeulerLagrange} only needs to be solved for one component,
say for $\sigma = \alpha$.
For unrestricted \HF, on the other hand,
this restriction is not applied~\cite{Pople1954}
and both components may diverge during an \SCF.

The spin component restriction of \RHF
implies that only closed-shell atomic systems can be treated.
For open-shell atoms typically \UHF is employed instead.
As already discussed in section \ref{sec:ImportanceAM}
an \UHF treatment of open-shell atoms, however,
typically suffers from issues related to a breaking
of spherical symmetry.

To illustrate this, consider the carbon atom with its nine-fold degenerate
$^3P$ ground state.
Not considering the spin degeneracy,
three energetically equivalent ground state Slater determinants exist,
which differ only in projected angular momentum $L_z$.
In a full configuration interaction treatment,
spherical symmetry could therefore be recovered
forming the ground state wave function
from a linear combination of these determinants.
For \UHF this is not possible due to
the single-determinant nature of \HF.
As a result the \UHF density matrix is symmetry-broken
and the \SCF procedure yields orbitals that are no longer of pure
$s$, $p$, $d$, \ldots \,character.
Such issues are naturally not restricted to ground states with a $P$ term,
but will occur similarly for all atoms with a ground state
of total angular momentum $L > 1$.

An additional approximation
to circumvent this behaviour is to employ
\textit{fractional occupation numbers}~(FON).
This approach emerged from developments to reproduce the
spectra of radical hydrocarbon species~%
\cite{LonguetHiggins1955,Dewar1968,Slater1969,Dewar1970,Ellison1971},
where the so-called \textit{half-electron method}~\cite{Carsky1972} was suggested
as a simpler alternative to the restricted open shell~\cite{Roothaan1960} procedure.
In the context of \UHF the FON approach distributes the
valance electrons evenly
over those valence orbitals differing only in the magnetic quantum number.
For the open-shell atoms of the second and third
period considered in this work,
this implies that an equal electron population
in the $2p$ or $3p$ orbitals are achieved
by selecting fractional values between $0$ and $1$
for those entries of the occupation matrix $\mat{f}^\sigma$
corresponding to said orbitals.
This effectively allows the \UHF procedure to converge to
a determinant, which is an average over those $2L+1$ degenerate
determinants one would actually need to combine in order to
recover spherical symmetry.

It should be noted, however, that a fractional occupation is no longer in
accordance with the Aufbau principle,
where the entries of $\mat{f}$ would be either $1$,
namely for all occupied orbitals,
or $0$, for all virtual orbitals.
This implies (1) that the resulting \HF density matrices are
no longer idempotent and (2) that the obtained solution
of Eq.~\eqref{eqn:SCFeulerLagrange}
cannot be a stationary point of the \HF minimisation problem~%
\cite{Bach1994}.
In other words the FON approach represents an additional approximation
on top of \UHF and the obtained energies will be \emph{higher}
compared to integer occupation.
As outlined in Ref. \onlinecite{Dewar1968}
with respect to the half-electron methods, however,
the difference between the integer and fractional approaches
can be expected to be small,
such that for many practical calculations both methods are typically
similarly suitable.

\subsection{Probing the required maximal angular momentum in \HF calculations}
\label{sec:RMSOl}
Summarising the discussion in section \ref{sec:BasisCS},
it is clear that choosing a suitable, but small value for $\lmax$,
to reach the desired level of accuracy
is important for \CS-based discretisations, too.
Similarly, understanding the angular momentum quantum
numbers required in a discretisation basis
can help to understand the properties of
quantum-chemical methods.

From an intuitive point of view
one would not expect all angular momentum to be equally important
for the description of the electronic ground state of a particular atom.
In beryllium, for example, only the $1s$ and $2s$ atomic orbitals are occupied,
such that only angular momentum $l = 0$ seems to be required.
Conversely all \CS functions with $l > 0$ should
contribute only very little, if at all.
Guided by this hypothesis the
\textit{root mean square occupied coefficient} per angular momentum $l$~($\RMSOl$),
is defined as
\begin{equation}
\RMSOl =
	\sqrt{
	\sum_{\substack{(n,l,m)\\ \in \Ibas}}
	\sum_{i} \sum_{\sigma\in\{\alpha, \beta\}}
	\frac{1}{N_\text{elec}^\sigma \ N_{\text{bas}, l}}
	\Big(C^\sigma_{\mu, i} \, f^\sigma_{ii} \Big)^2
	},
	\label{eqn:DefRMSOl}
\end{equation}
where $i$ runs over all \SCF orbitals,
$\Ibas \subset \Itpf$ is the selected set of
index triples $\mu \equiv (n,l,m)$ for the \CS basis functions
and $N_\text{elec}^\sigma$ are the number of electrons of spin $\sigma$.
Furthermore
$C^\alpha_{\mu i}$, $C^\beta_{\mu i}$, $f^\alpha_{ii}$
and $f^\beta_{ii}$ are the matrix
elements of the orbital coefficient matrices and occupation matrices
introduced in section \ref{sec:HFfractional}
and
\begin{equation}
	N_{\text{bas}, l} := \Big| \left\{ (n',l',m') \, \big|\, (n',l',m') \in \Ibas
		\ \text{and} \ l' = l \right\} \Big|
	\label{eqn:Nbasl}
\end{equation}
is the number of basis functions in the \CS basis which have angular momentum
quantum number $l$.
By construction $\RMSOl$ is the root mean square~(\RMS)
coefficient for a particular angular
momentum quantum number $l$ in the occupied SCF orbitals.
It therefore provides a measure,
which angular momentum quantum numbers $l$
of the current basis set are used in a significant amount
for describing the \HF wave function
--- namely those where $\RMSOl$
is above the convergence threshold of the \SCF procedure.

To use this quantity for finding a good value of $\lmax$
to restrict the \CS basis first
a \HF calculation is performed,
where the employed \CS basis set is only restricted by $\nmax$.
This value $\nmax$ should be chosen carefully,
since on the one hand too large a value leads to rather large basis sets
and thus potentially expensive calculations
and on the other hand too small a value
implies that $l$ does not reach large enough values to observe a visible trend.
In this work we used a $(6,5,5)$ basis set for this step.
Afterwards an $\RMSOl$ plot,
\ie plot $\RMSOl$ versus $l$, is produced and the trends observed.
Since larger $l$ implies more angular nodes
thus higher kinetic energy,
larger values of $l$ will become less and less significant,
\ie $\RMSOl$ will decrease.
Inspecting the plot an $\lmax$ can then be chosen
such that those angular momentum quantum numbers larger $\lmax$
can be considered insignificant.
See section \ref{sec:ConvSize} for examples.

In analogy to Eq.~\eqref{eqn:DefRMSOl}
we can furthermore compute a root mean square coefficient value per
basis function angular momentum quantum number $l$
for each \SCF orbital.
This quantity will be used in section \ref{sec:ConvSize}
to explain the behaviour of $\RMSOl$ plots.

\subsection{Computational details and reference values}
\label{sec:ComputDetails}
All \HF computations presented in this work
were obtained using the \sturmint~\cite{sturmintWeb}
Coulomb Sturmian integral library
in combination with the \molsturm framework~\cite{molsturmDesign}
or the \texttt{SelfConsistentField.jl}~\cite{SelfConsistentFieldJl}
code to drive the \SCF computation.
Postprocessing and plotting was done
in \python~\cite{Python} and \julia~\cite{Julia}
using \molsturm, numpy~\cite{Walt2011,scipyWeb},
pandas~\cite{pandasWeb} and matplotlib~\cite{Matplotlib}.

\RHF was employed to compute the energies of closed-shell atoms,
whereas \UHF was used for open-shell systems.
If not explicitly mentioned otherwise,
the \UHF results refer to calculations employing
integer occupation numbers.
For carbon and oxygen some \UHF with fractional occupation were underdone
as well, where an occupation of $2/3$ was used for the $2p\alpha$
orbitals of carbon and of $1/3$ for the $2p\beta$ orbitals of oxygen.

\ctable[
    cap=Hartree-Fock reference results used for comparison,
    caption=Reference values used for comparison of the \CS-based results and for estimating errors in the \CS values. The CBS extrapolation was done following \citet{Jensen2005}. ,
    botcap,
    doinside=\smaller,
    mincapwidth=0.98\columnwidth,
    footerwidth=0.88\columnwidth,
    label=tab:HFReference,
]{lr@{}l@{\hspace{40pt}}lr@{}l}{
    \tnote[U]{unrestricted HF with integer occupation numbers}
    \tnote[F]{unrestricted HF with fractional occupation numbers}
    \tnote[R]{restricted HF}
    \tnote[a]{CBS extrapolation using cc-pVDZ to cc-pV5Z~\cite{Dunning1989,Woon1993}}
    \tnote[b]{Values taken from \citet{Morgon1997}}
    \tnote[c]{CBS extrapolation using cc-pVTZ to cc-pV6Z~\cite{Dunning1989,Woon1993,Wilson1996,VanMourik2000,Prascher2011}}
}{
    \FL
    system & \multicolumn{2}{c}{$E_\text{HF}$} & system & \multicolumn{2}{c}{$E_\text{HF}$} \ML
    Li & $-7$ & $.4327376$\tmark[a,U] & Na & $-161$ & $.8589459$\tmark[a,U] \NN
    Be & $-14$ & $.57302317$\tmark[b,R] & Mg & $-199$ & $.61463642$\tmark[b,R] \NN
    B & $-24$ & $.5334831$\tmark[a,U] & Al & $-241$ & $.8808503$\tmark[c,U] \NN
    C & $-37$ & $.6937751$\tmark[c,U] & Si & $-288$ & $.8589476$\tmark[c,U] \NN
    C & $-37$ & $.5313456$\tmark[c,F] &  &  &  \NN
    N & $-54$ & $.4046409$\tmark[c,U] & P & $-340$ & $.7192829$\tmark[c,U] \NN
    O & $-74$ & $.8192096$\tmark[c,U] & S & $-397$ & $.5133666$\tmark[c,U] \NN
    O & $-74$ & $.624862$\tmark[c,F] &  &  &  \NN
    F & $-99$ & $.4166858$\tmark[c,U] & Cl & $-459$ & $.4899302$\tmark[c,U] \NN
    Ne & $-128$ & $.54709811$\tmark[b,R] & Ar & $-526$ & $.8175128$\tmark[b,R] \LL
}
The estimation of errors and convergence
was done by comparing our \CS-based \HF results
with the reference values of Table \ref{tab:HFReference}.
These include, for \RHF calculations,
the numerical \RHF energies obtained by \citet{Morgon1997}.
For \UHF calculations the complete basis set~(\CBS) limit
was extrapolated following the approach of \citet{Jensen2005}
applied to calculations with the Dunning cc-pV$n$Z family
of \cGTO basis sets~\cite{Dunning1989,Woon1993,Wilson1996,VanMourik2000,Prascher2011}.

\section{Results and discussion}
\label{sec:Results}

This section presents the results of our convergence study
of \CS basis sets for discretising the \HF problem,
obtained for the atoms of
the second and third period of the periodic table.
We expect the outlined procedures to be of general character,
however,
such that they could be applied
to the remainder of the periodic table as well.

\subsection{Convergence with respect to basis set size
and angular momentum analysis}
\label{sec:ConvSize}

As described in section \ref{sec:BasisCS}
the construction of Coulomb Sturmian basis sets
consists of the selection of roughly two types of parameters.
Firstly $\nmax$ and $\lmax$,
which fix the size of the basis
and secondly $\kexp$, which communally
fixes the exponential falloff of all basis functions.

\begin{figure}
	\centering
	\includegraphics[scale=0.8]{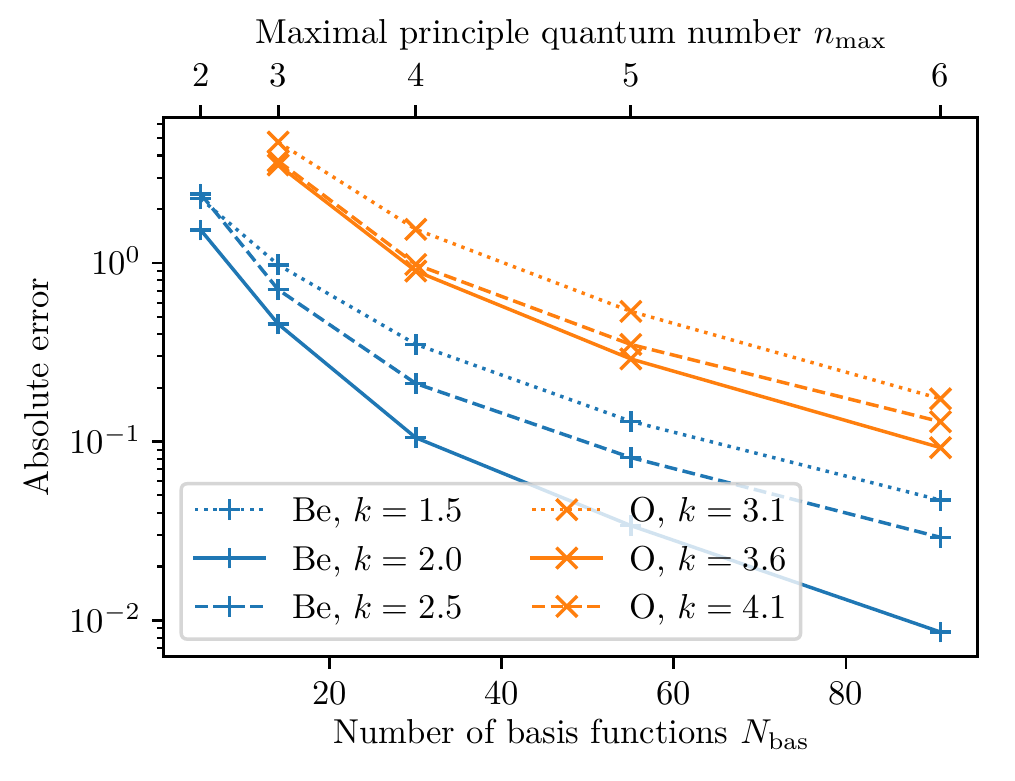}
	\caption
	{
		Absolute error in the \HF energy versus number of basis functions
		in a \CS basis set only restricted by the maximal
		principle quantum number $\nmax$ (top axis).
		The blue curves are \RHF calculations
		of beryllium, the orange curves
		\UHF calculations of oxygen, each with different \CS exponents $\kexp$.
		The reference values for the error computation
		were taken from Table \ref{tab:HFReference}.
}
	\label{fig:ErrorHF_vs_shell_k}
\end{figure}

In this section we will primarily
discuss convergence with respect to the first aspect,
\ie the \CS basis set size.
As outlined in the previous sections,
both the completeness of the \CS radial part $R_{nl}$
as well as the \CS functions $\varphi_{nlm}$
is independent of the exponent $\kexp$.
The general convergence trend with respect
to increasing basis set size can thus be
expected to be independent of the value of $\kexp$ as well.
There is, however, a notable effect
on the convergence rate with increasing basis size.
This is demonstrated in Fig.~\ref{fig:ErrorHF_vs_shell_k},
which shows the convergence of the \HF energies
of beryllium and oxygen, respectively.
In both cases the absolute error versus the reference values%
~(see Table \ref{tab:HFReference})
is plotted against the size of the employed \CS basis sets,
which are only restricted by
the indicated maximal principle quantum number $\nmax$.
As expected by the Courant-Fischer variational theorem~\cite{Helffer2013}
a decrease in error can be observed for
larger basis sets, \ie larger values for $\nmax$.
For all curves the convergence appears to be sub-linear
with the best rate of convergence achieved for $\kexp = 2.0$ for beryllium
and $\kexp = 3.6$ for oxygen.
Larger as well as smaller exponents worsen the convergence rate,
which will be discussed in more detail in section \ref{sec:ConvK}.
For the discussion in this section
it is sufficient to note,
that convergence is achieved regardless of the value of $\kexp$,
but some optimal, atom-dependent value exists for each \CS basis set,
where the \HF energy is lowest.

Such observations are in agreement with previous results
obtained by \citet{Avery2017},
where they approximated Slater-type orbitals~(STO)
in a basis of Coulomb Sturmians.
In their treatment they also found that convergence
is faster the closer the \CS exponent of the basis
to the STO exponent of the function to be approximated,
but convergence occurred in either case.
For understanding the convergence behaviour with respect to
increasing the basis set size
the dependency on $\kexp$ can thus be largely ignored
--- provided that for each atom a reasonable value for $\kexp$ is chosen.

\begin{figure}
	\centering
	\includegraphics[scale=0.8]{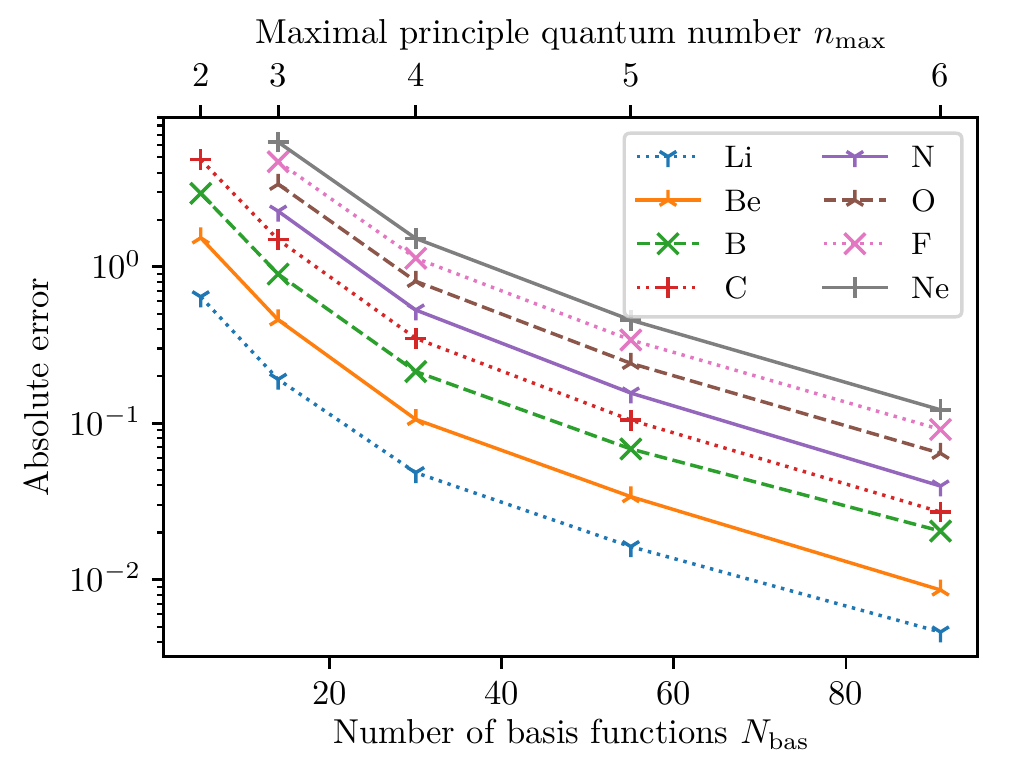}
	\caption
	{
		Plot of the absolute error in the \HF energy versus the number of basis
		functions in a \CS basis restricted by $\nmax$.
		For the closed-shell atoms Be and Ne
		the restricted \HF procedure was used,
		whereas for the other systems \UHF with integer occupation was employed.
		For each calculation of a particular atom the same value of $\kexp$ was used,
		which was taken within $0.01$ to the optimal exponent
		of this atom at $(6,5,5)$ level.
		The errors were computed against the reference
		values from table~\ref{tab:HFReference}.
}
	\label{fig:ErrorHF_vs_shell}
\end{figure}
The convergence trend observed in Fig.~\ref{fig:ErrorHF_vs_shell_k}
for beryllium and oxygen appears to be more general.
Our investigations show that it can
at least be replicated in a similar fashion
for the other atoms of the second period~(see Fig.~\ref{fig:ErrorHF_vs_shell})
as well as the third period.
In all these calculations the convergence is noticeably sublinear
and overall comparatively slow.
Already for the second half of the second period
reaching below absolute errors of $0.1$ Hartree
requires beyond $80$ basis functions,
making calculations with Coulomb Sturmian basis sets
only restricted by $\nmax$ rather impractical.

\begin{figure}
	\centering
	\includegraphics[scale=0.8]{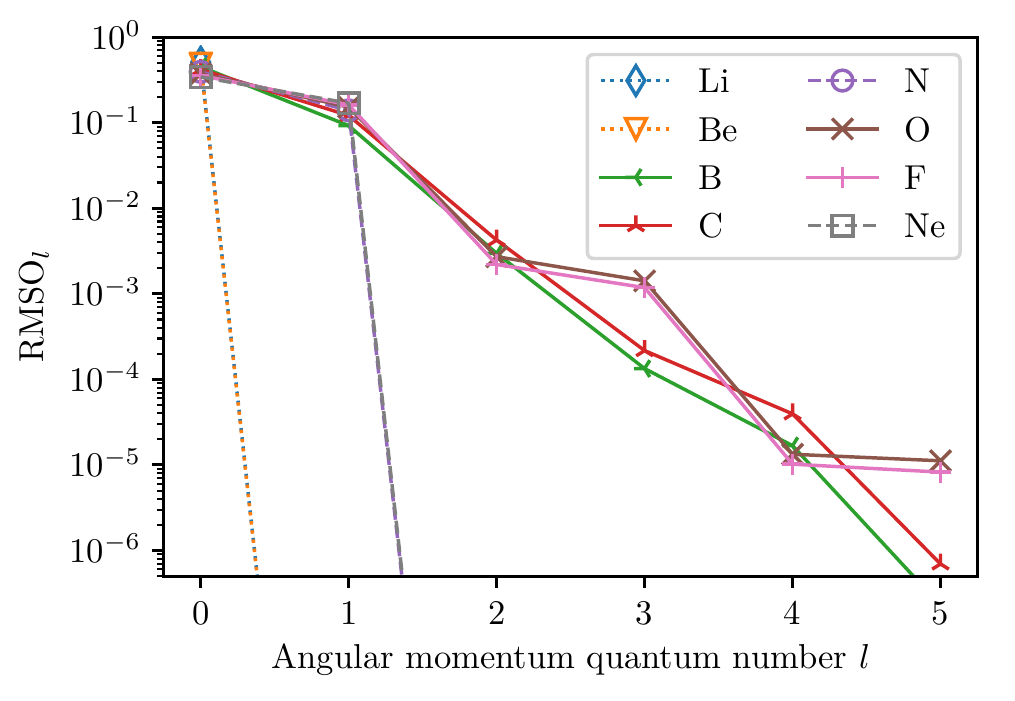}
	\caption
	{
		Plot $\RMSOl$ vs. $l$ for the \HF ground state
		of the atoms of the second period
		if a $(6,5,5)$ \CS basis is employed.
		In each case $\kexp$ was
		taken within $0.01$ to the optimal exponent.
		For \ce{Be} and \ce{Ne} a \RHF procedure
		was used, for the other cases \UHF with integer
		occupation numbers.
		The graphs for \ce{Li} and \ce{Be}
		as well as \ce{N} and \ce{Ne}
		with their respective sharp drop features
		are almost superimposed.
	}
	\label{fig:RMSOl_period2}
\end{figure}

\begin{figure}
	\centering
	\includegraphics[scale=0.8]{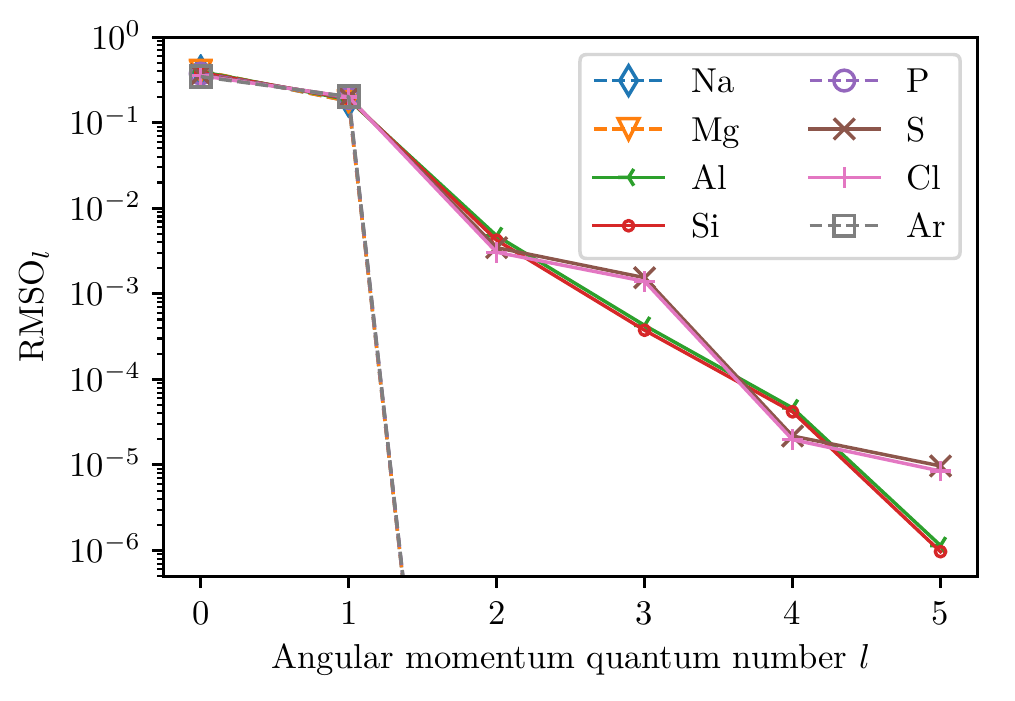}
	\caption
	{
		Plot $\RMSOl$ vs. $l$ for the \HF ground state
		of the atoms of the third period
		if a $(6,5,5)$ \CS basis is employed.
		In each case $\kexp$ was taken close to the optimal value.
		For \ce{Mg} and \ce{Ar} a \RHF procedure
		was used, for the other cases \UHF with integer
		occupation numbers.
		The graphs for \ce{Na}, \ce{Mg}, \ce{P}
		and \ce{Ar}
		are almost superimposed.
	}
	\label{fig:RMSOl_period3}
\end{figure}

In section \ref{sec:BasisCS} we deduced
that the basis size scaling with respect to $\nmax$
can be reduced from cubic to linear if
the basis can be restricted by $\lmax$ as well,
which evidently has an impact on the convergence speed.
In order to find bounds for $\lmax$,
the $\RMSOl$ measure introduced in section \ref{sec:RMSOl}
is applied
to the \SCF coefficients obtained in a $(6,5,5)$ \CS basis.
\begin{figure}
	\centering
	\includegraphics[scale=0.8]{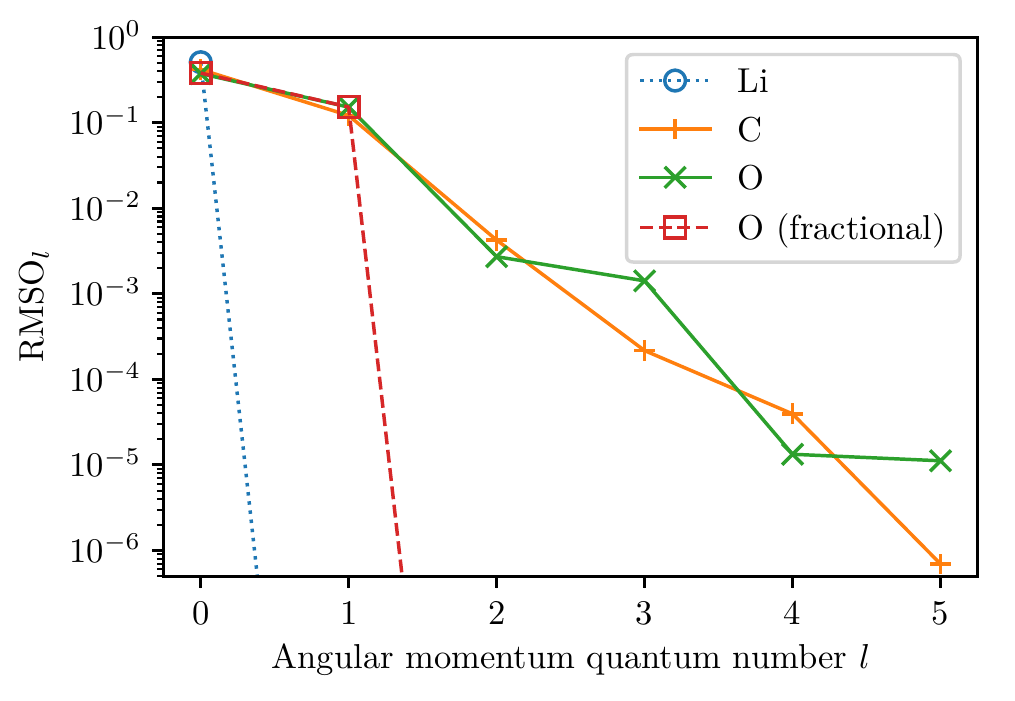}
	\caption
	{
		Plot $\RMSOl$ vs. $l$ for the \UHF ground state
		of the atoms of the second period
		if a $(6,5,5)$ \CS basis is employed.
		In each case $\kexp$ was
		taken within $0.01$ to the optimal exponent.
		For oxygen both a case with integer and a case with
		fractional occupation numbers is depicted.
	}
	\label{fig:RMSOl_period2_summary}
\end{figure}
For the \UHF calculations of lithium, carbon and oxygen,
plots of $\RMSOl$ vs. $l$ are shown
in Fig.~\ref{fig:RMSOl_period2_summary}.
Corresponding plots for the other atoms of the second
and third period can be found
in Figures~\ref{fig:RMSOl_period2} and~\ref{fig:RMSOl_period3}.

In all aforementioned plots roughly two trends can be identified.
The first is a very pronounced drop in $\RMSOl$,
which occurs once a particular
angular momentum value $l$ has been reached.
For example, in Fig.~\ref{fig:RMSOl_period2_summary} this is observed
for lithium (between $l = 0$ and $1$)
and oxygen with fractional occupation numbers (between $l = 1$ and $2$).
The second is a decreasing staircase pattern,
where the $\RMSOl$ value decreases only very moderately over
the range of considered angular momentum quantum numbers.
In Fig.~\ref{fig:RMSOl_period2_summary}, for example,
this is observed for
oxygen with integer occupation numbers as well as for carbon.
Considering the atoms of the second and the third period alltogether,
the rapid-drop-type $\RMSOl$ plots
are obtained for those atoms with an $S$ ground state term.
That is those, which are either closed-shell
like \ce{Be}, \ce{Ne}, \ce{Mg} or \ce{Ar}
or which have a half-filled $s$- or $p$-shell
like \ce{Li}, \ce{N}, \ce{Na} or \ce{P}.
In these cases the drop occurs exactly
where one would expect by looking
at the largest angular momentum of
the occupied atomic orbitals,
\ie between $l=0$ and $l=1$ for \ce{Li} and \ce{Be},
and between $l=1$ and $l=2$ for the other mentioned cases.
In contrast to this the atoms
with a $P$ ground state term, namely
\ce{B}, \ce{C}, \ce{O}, \ce{F}, \ce{Al}, \ce{Si}, \ce{S} and \ce{Cl},
follow the decreasing staircase pattern,
but only if fractional occupation numbers are not used.

\begin{figure}
	\centering
	\includegraphics[scale=0.8]{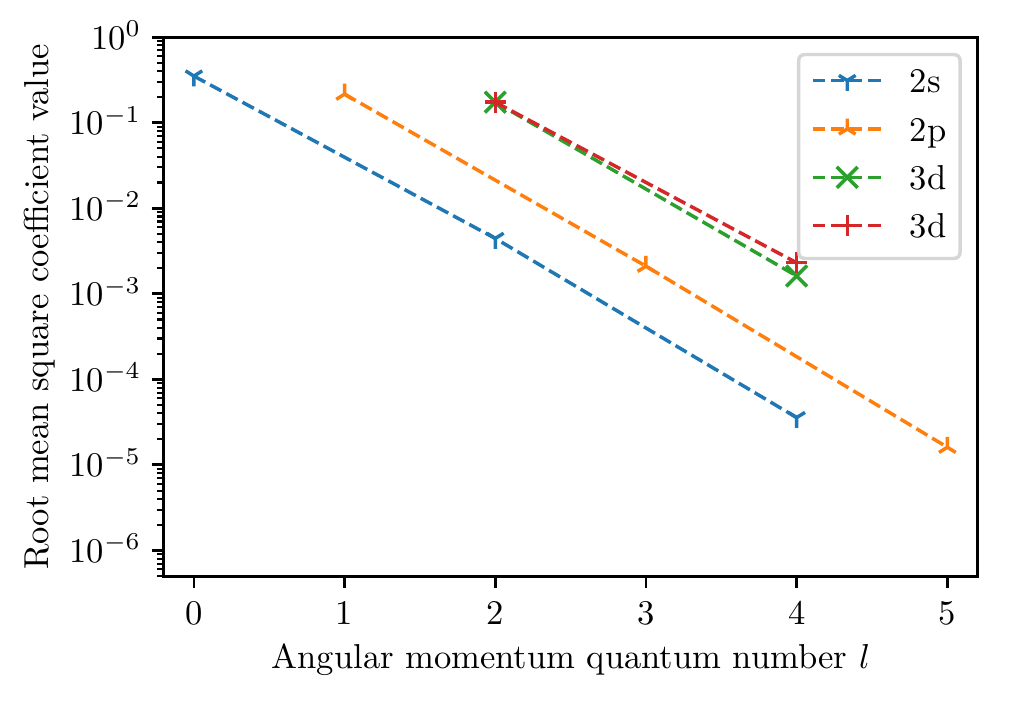}
	\caption{
		Root mean square coefficient value per
		basis function angular momentum quantum number $l$
		for selected orbitals of oxygen.
		The atom is modelled
		in a $(6,5,5)$ \CS basis using \UHF.
	}
	\label{fig:RMSLF_O}
\end{figure}
\begin{figure}
	\centering
	\includegraphics[scale=0.8]{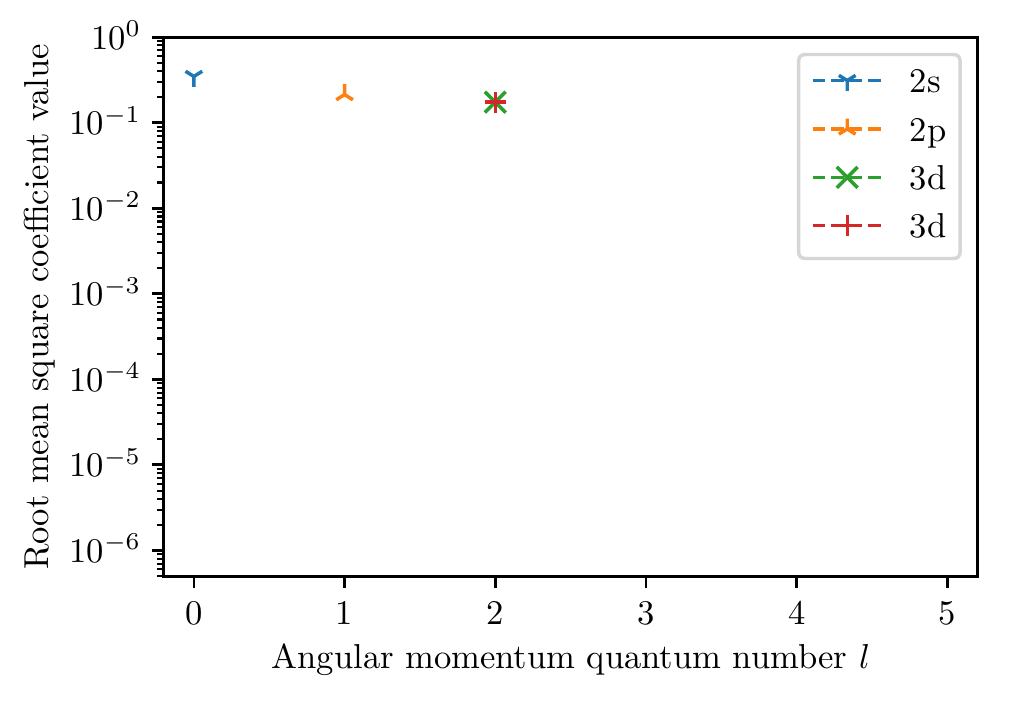}
	\caption
	{
		Root mean square coefficient value per
		basis function angular momentum quantum number $l$
		for selected orbitals of nitrogen.
		The atom is modelled
		in a $(6,5,5)$ \CS basis using \UHF.
	}
	\label{fig:RMSLF_N}
\end{figure}
\begin{figure}
	\centering
	\includegraphics[scale=0.8]{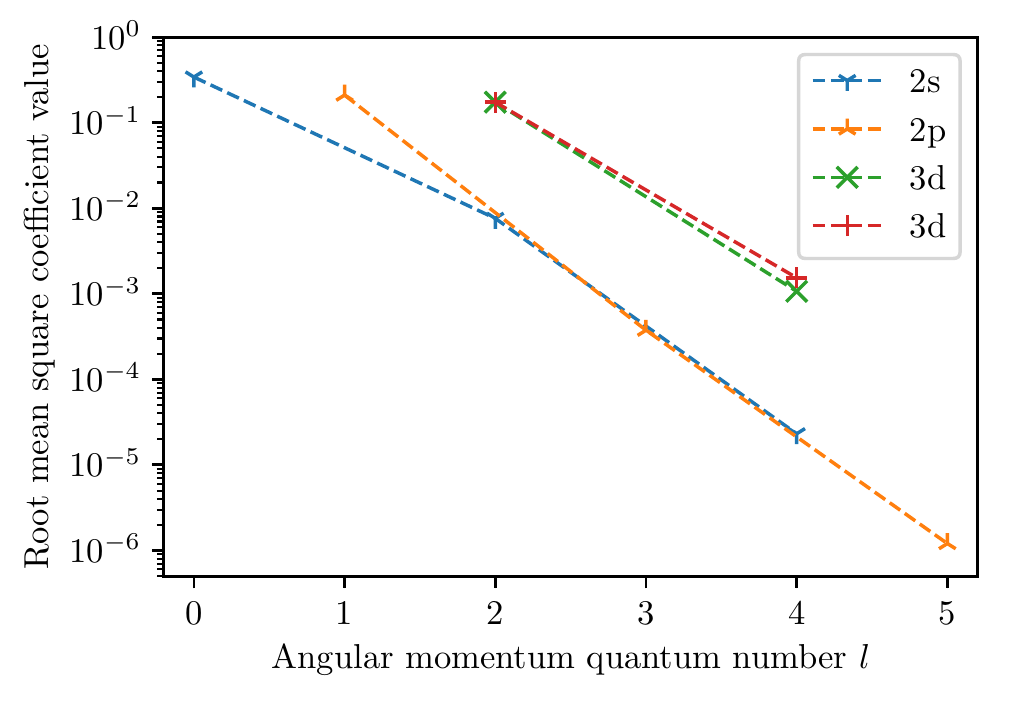}
	\caption
	{
		Root mean square coefficient value per
		basis function angular momentum quantum number $l$
		for selected orbitals of carbon.
		The atom is modelled
		in a $(6,5,5)$ \CS basis using \UHF.
	}
	\label{fig:RMSLF_C}
\end{figure}

A hint to explain the second type of behaviour is obtained
by looking at the \RMS coefficient values
per basis function angular momentum quantum number $l$ for each orbital%
~(for details see section~\ref{sec:RMSOl}).
A plot of these values against $l$ is shown
in Fig.~\ref{fig:RMSLF_O} for oxygen.
Surprisingly, the $2s$ \UHF orbital not only
consists of basis functions with $l=0$,
but furthermore of functions with $l=2$ and $l=4$
in the employed $(6,5,5)$ \CS basis.
Similar observations can be made for the $2p$ and $3d$ functions,
which are not angular-momentum-pure any longer,
but consist of angular momentum in steps of $2$ apart.
This behaviour explains why $\RMSOl$ plots for oxygen
does not show the expected drop from $l=1$ to $l=2$,
since the higher angular momenta play a role
for the occupied $s$-type and $p$-type \SCF orbitals as well.
This observed behaviour is in perfect agreement with the breaking of spherical
symmetry previously observed in \UHF calculations%
~\cite{Cook1981,Fukutome1981,Cook1984,McWeeny1985}.
As described in Ref. \onlinecite{McWeeny1985} the \UHF wave function
in such cases is not spherically, but \emph{axially} symmetric.
On the level of the \SCF orbitals themselves,
this is realised by mixing with higher angular momentum basis functions.
For illustration consider amending
a spherically symmetric $s$ orbital with
a fraction of a $d_{z^2}$ basis function $d_{z^2}$.
This causes a stretching of the orbital along the $z$ axis,
which makes it axially symmetric.
Similarly the $p_x$, $p_y$ and $p_z$ orbitals
may be amended with $f_{xz^2}$, $f_{yz^2}$ and $f_{z^3}$
to elongate them in $z$ direction.
Since parity may not be violated,
an orbital may only consist of basis functions with either even
or with odd angular momentum,
explaining the pattern of Fig.~\ref{fig:RMSLF_O}.
For other calculations,
which show the decreasing staircase $\RMSOl$ pattern,
similar plots to Fig.~\ref{fig:RMSLF_O}
with smeared-out angular momentum are obtained.
See for example carbon in Fig.~\ref{fig:RMSLF_C}.
Conversely, if the \SCF orbitals are pure in angular momentum,
a clear drop in the $\RMSOl$ plots is observed.
One example is nitrogen~(Fig.~\ref{fig:RMSLF_N})
or oxygen at \UHF level with fractional occupations.
This confirms our discussion in section \ref{sec:HFfractional}
indicating that an $S$ ground state term
or an \UHF treatment with fractional occupation numbers
prevents a breaking of spherical symmetry.

\begin{figure}
	\centering
	\includegraphics[scale=.8]{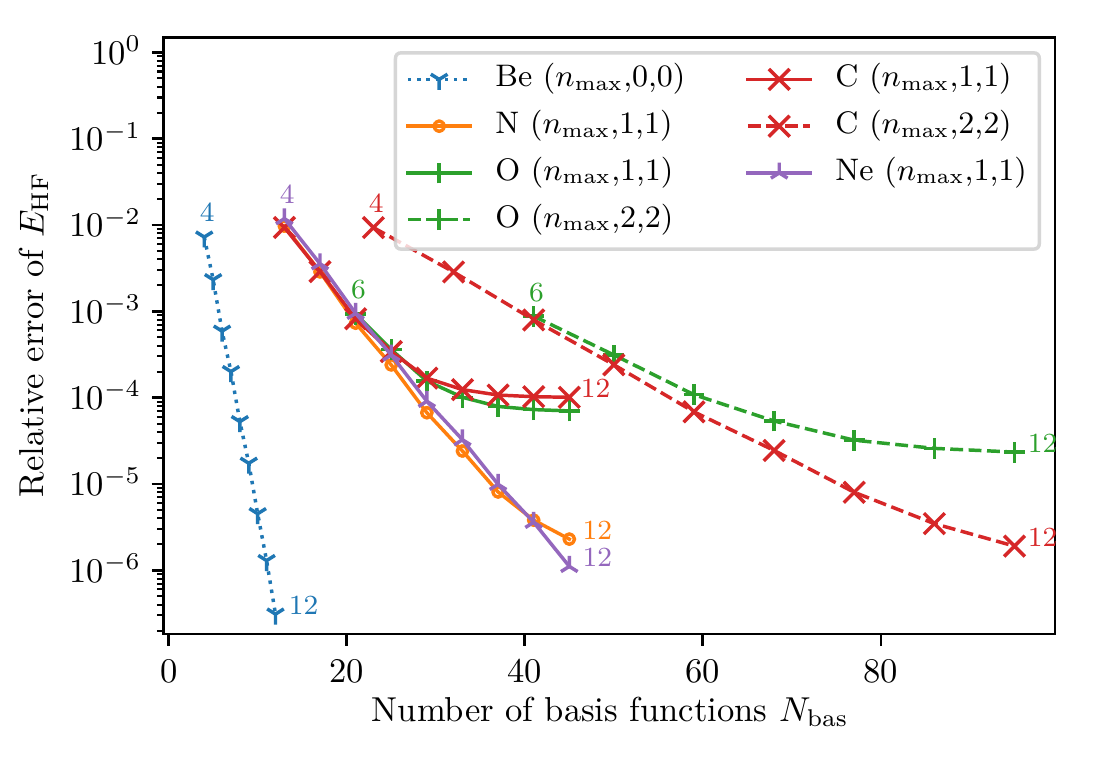}
	\caption{
		Relative error in $E_\text{HF}$ versus the number of basis functions
		for selected \CS basis sets of the form $(\nmax, \lmax, \lmax)$.
		The connected points show basis set progressions
		in which the maximum principle quantum number 
		$\nmax$ is increased in steps, while $\lmax$ is kept fixed.
		The first and last value for $\nmax$ are denoted as small numbers
		next to appropriate datapoints.
		The same line type is used for all progressions of the same $\lmax$
		and the same colour and marker for all progressions of the same atom.
		For \ce{Be} and \ce{Ne} \RHF was used and
		for \ce{N}, \ce{C} and \ce{O} \UHF.
	}
	\label{fig:ErrorHF_vs_nlm}
\end{figure}
\begin{figure}
	\centering
	\includegraphics[scale=.8]{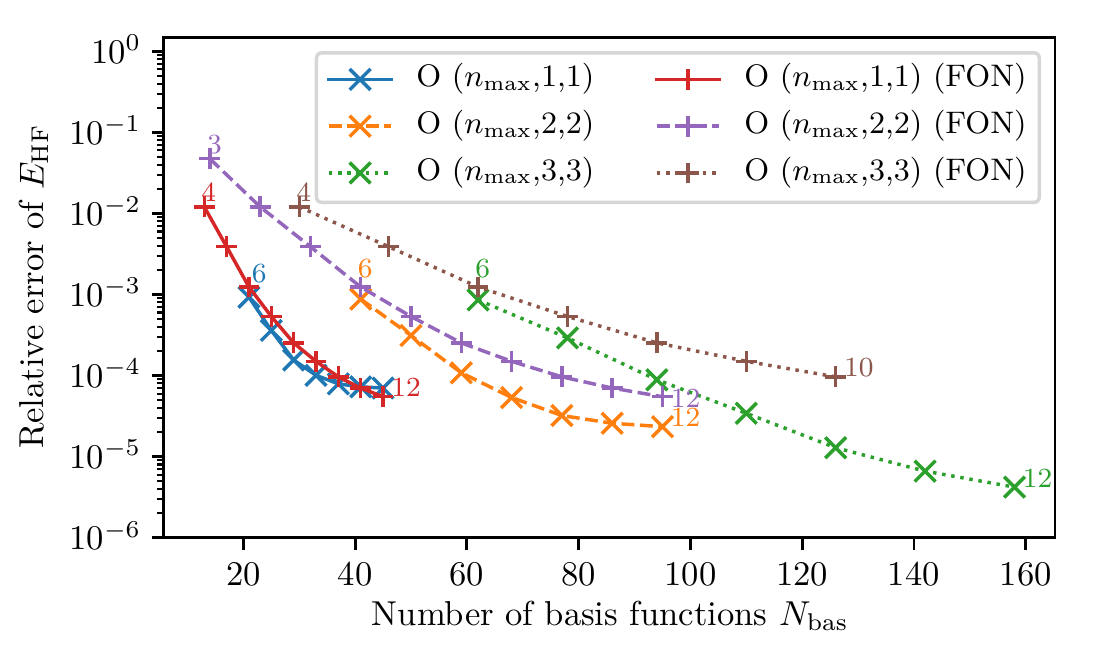}
	\caption{
		Relative error in $E_\text{HF}$ versus the
		number of basis functions for selected basis set progressions.
		Shown are \UHF calculations
		of oxygen using either integer or fractional
		occupation numbers~(FON).
		The same display conventions as Fig.~\ref{fig:ErrorHF_vs_nlm}
		are used.
	}
	\label{fig:ErrorHF_vs_nlm_O}
\end{figure}
With this in mind, a bound for $\lmax$ is easy to choose
for those $\RMSOl$ plots with a clearly observable drop,
namely exactly the value for the angular momentum quantum number
$l$ \emph{before} the drop is encountered.
For the cases with a decreasing staircase pattern
the selection is not so straightforward,
since $\lmax$ both influences the prefactor in the scaling
of basis size versus $\nmax$ (see Eq.~\eqref{eqn:NbasAmLimitedBasis})
as well as the angular discretisation a basis provides.
To observe the influence of different choices for $\nmax$ and $\lmax$
Fig.~\ref{fig:ErrorHF_vs_nlm} shows
\RHF or \UHF results obtained
using progressions of \CS basis sets, where $\lmax$ is limited
to either $0$, $1$ or $2$,
but $\nmax$ is ranged between $4$ and $12$.
In each case the relative error of the \HF energy
with respect to the reference values in Table \ref{tab:HFReference}
are plotted against the size of the \CS basis.
Those error values corresponding to the same atom and the same $\lmax$,
but different $\nmax$, are connected by lines.
In the following we will refer to such a sequence of calculations,
in which only $\nmax$ differs by the term \textit{progression}.
In all calculations for a particular atom the \CS exponent
$\kexp$ has been kept constant.

\begin{figure}
	\centering
	\includegraphics[scale=0.8]{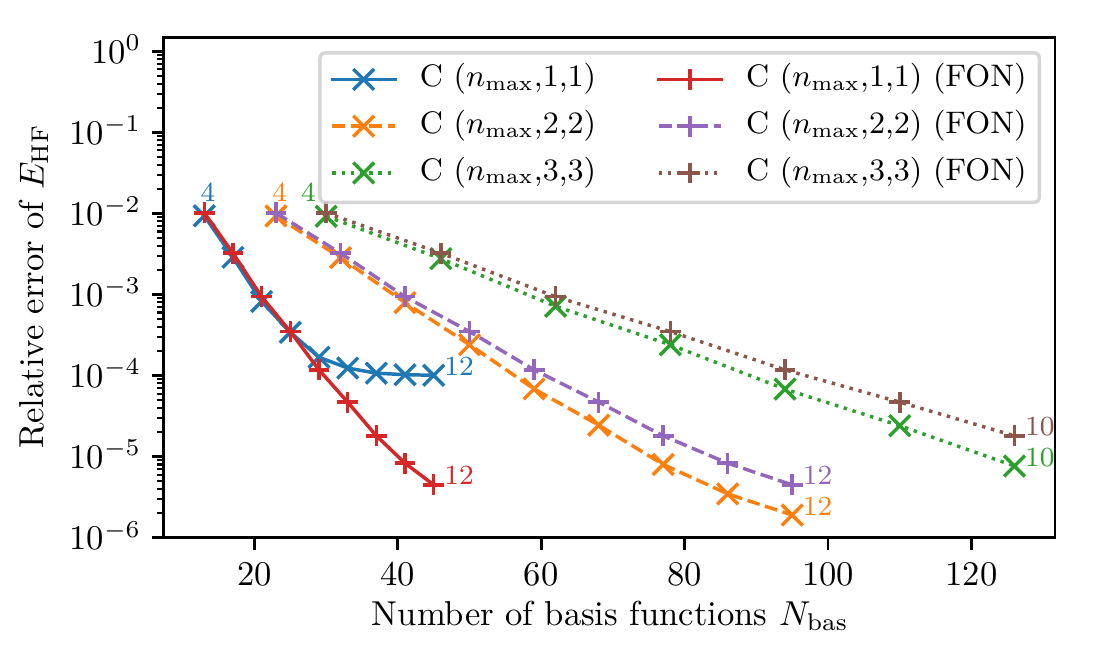}
	\caption{
		Relative error in $E_\text{HF}$ versus the
		number of basis functions for selected basis set progressions.
		Shown are \UHF calculations
		of carbon using either integer or fractional
		occupation numbers~(FON).
		The same display conventions as figure \ref{fig:ErrorHF_vs_nlm}
		are used.
	}
	\label{fig:ErrorHF_vs_nlm_C}
\end{figure}

In agreement with the conclusions from the $\RMSOl$ plots
a very good convergence with respect to increasing
$\nmax$ is observed for beryllium, nitrogen and neon
even if the angular momentum is restricted by
$\lmax = 0$ or $\lmax = 1$.
A further increase of $\lmax$ does not improve
the obtained error regardless of the value of $\nmax$.
Since the basis now grows faster as $\nmax$ increases,
the convergence rate is slower in such cases, however.
This is in contrast to oxygen and carbon.
As the $\RMSOl$ plots in Fig.~\ref{fig:RMSOl_period2_summary} suggest,
the angular momentum values $l > 2$ are required
for a proper description of the symmetry broken \UHF ground state.
It is therefore no surprise that the convergence of the \HF energy
begins to stagnate for the $\nmax$-progressions
with $\lmax = 1$ as well as $\lmax = 2$.
In these cases a relevant part of the ground state wave function
cannot be represented in the available angular discretisation
and at some point the resulting error in the angular discretisation dominates.
Improving the radial discretisation by increasing $\nmax$
thus cannot decrease the net error any further.
The obtained limiting relative error depends
on $\lmax$ --- with larger values of $\lmax$ allowing
a basis progression to yield a lower error limit.
The obtained limits are further system-dependent and their trends with $\lmax$
can be understood looking at the
decreasing staircase patterns.
For example, the limiting error in the $\lmax=1$ and $\lmax=2$ progressions
for oxygen is almost unchanged,
whereas it is significantly smaller for carbon.
At the same time,
considering the $\RMSOl$ plots in Fig.~\ref{fig:RMSOl_period2_summary},
the decrease in the $\RMSOl$ value
from $l=1$ to $l=2$ is small for \ce{O},
but a good order of magnitude for the \ce{C} atom.
Going to larger angular momentum
the $\RMSOl$ plot for oxygen with integer occupation
undergoes a significant decrease between $l=2$ and $l=3$, however.
In agreement the limiting error of a oxygen $\lmax=3$ progression decreases
as well, see Fig.~\ref{fig:ErrorHF_vs_nlm_O},
which shows the $\lmax=1$, $\lmax=2$ and $\lmax=3$
progressions for oxygen with both integer
and fractional occupations.
As discussed in section \ref{sec:HFfractional}
fractional occupation numbers prevent symmetry breaking,
such that pure angular momentum \SCF orbitals are obtained.
As a consequence only pure $s$ and $p$ functions
are occupied and no improvement
in \UHF energies are obtained for the progressions
with $\lmax > 1$.
For comparison an equivalent plot to Fig.~\ref{fig:ErrorHF_vs_nlm_O}
for carbon is shown in Fig.~\ref{fig:ErrorHF_vs_nlm_C}.

An interesting aspect to note
in all plots of relative error versus
basis set size is the initial convergence,
which seems to follow a linear behaviour in all depicted cases.
Furthermore the initial rate appears to depend only on $\lmax$,
but notably not on the system under investigation.
As the progression continues, most curves bent off to become sublinear.
A closer inspection, however,
reveals two kinds of trends,
which are best visible for the $\lmax=1$ progressions
in Fig.~\ref{fig:ErrorHF_vs_nlm_O}
as well as Fig.~\ref{fig:ErrorHF_vs_nlm_C}.
For the integer occupation numbers,
the previously discussed stagnation of convergence is observed,
which we could explain with reference
to the decreasing staircase pattern
in the $\RMSOl$ plots and too small a value for $\lmax$.
For the fractional occupation numbers,
the curves do not not completely stagnate,
but merely slows down.
This is also observed for some other cases,
\eg \ce{N} in Fig.~\ref{fig:ErrorHF_vs_nlm},
where the $\RMSOl$ plot allows to point out  a particular value $\lmax$,
where all angular discretisation of the \HF wave function should be obtained.
Such sublinear convergence behaviour is not an unusual result
in electronic structure theory.
See for example Ref. \onlinecite{Jensen2005}
for a discussion of \CBS extrapolations using \cGTO basis
sets or Ref. \onlinecite{Bachmayr2014} for error
estimates for even-tempered Gaussian-type basis sets.
In combination with the apparently system-independent initial convergence,
this indicates that rigorous \CBS extrapolation techniques
are within reach for \CS basis sets as well.

\subsection{Convergence with respect to the Coulomb Sturmian exponent $k$}
\label{sec:ConvK}

Having discussed convergence with \CS basis set size in the previous sections,
we now turn our attention to
the \CS exponent $\kexp$.
In Fig.~\ref{fig:ErrorHF_vs_shell_k} of section \ref{sec:ConvSize}
we already noted the convergence rate of \CS discretisations
to depend on $\kexp$ with some values giving
faster and some slower convergence.
For constructing a basis,
which approximates the wave function best given a particular
\CS basis size, a suitable exponent $\kexp$ needs to be chosen as well.
This section will discuss the influence of altering
the \CS exponent $\kexp$ in more detail.

\begin{figure}
	\centering
	\includegraphics[scale=0.8]{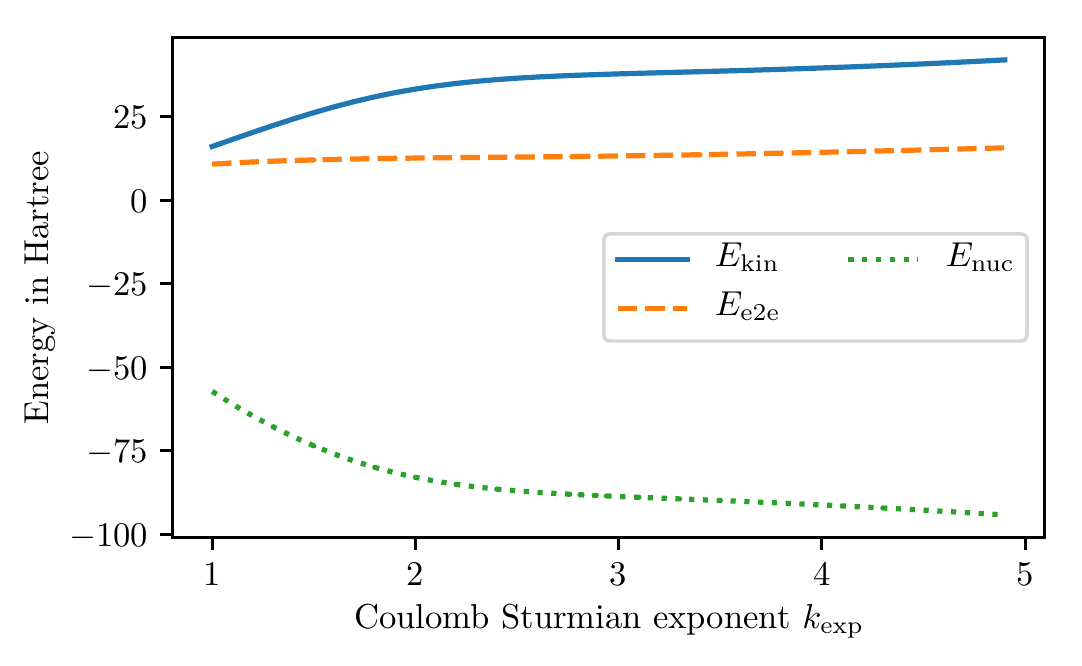}
	\caption
	{
		Plot of the \HF energy contributions
		of the carbon
		atom versus the Coulomb Sturmian exponent $\kexp$.
		All calculations are done in a $(5,2,2)$ \CS basis
		using \UHF.
	}
	\label{fig:EnergyTermsVsK}
\end{figure}

In the \CS basis functions,
$\kexp$ only occurs in the radial part, see Eq.~\eqref{eqn:CSdef}.
Through the exponential term $\exp(-\kexp r)$,
$\kexp$ influences how quickly the basis functions decay asymptotically
and in the form of a polynomial prefactor it determines
the curvature of the radial functions as they oscillate between the radial nodes.
Keeping this in mind let us consider Fig.~\ref{fig:EnergyTermsVsK},
which shows the changes to individual energy contributions
of the \HF ground-state energy as $\kexp$ is altered.
The largest changes are apparent for the nuclear attraction energy~($E_\text{nuc}$),
which decreases --- initially rather steeply --- as $\kexp$ is increased.
This can be easily understood from a physical point of view.
Since larger values of $\kexp$ imply a more rapid decay
of the basis functions,
the electron density on average stays closer to the nucleus,
which in turn leads to a lower (more negative) interaction energy
between electrons and nucleus.
The converse effect happens for smaller values of $\kexp$,
where the electron density is more expanded
and thus on average at larger distance from the nucleus.
In contrast the kinetic energy~($E_\text{kin}$)
is related to the curvature of the wave function,
which --- as described above --- increases for larger $\kexp$.
In other words the trends of nuclear attraction energy
and electronic kinetic energy
oppose each other,
with the kinetic energy being effected to a lesser degree.
On the scale depicted in Fig.~\ref{fig:EnergyTermsVsK}
the variation of the electron-electron interaction~($E_\text{e2e}$),
\ie both classical Coulomb repulsion as well as the exchange interaction combined,
is much less pronounced.
Only a very minor increase with $\kexp$ can be observed.
The physical mechanism is again similar to the nuclear attraction
energy term,
namely that larger $\kexp$ compresses the wave function
and thus lets the electrons reside more closely to another,
which increases the Coulomb repulsion between them.
The exchange interaction is effected as well,
but changes are smaller and thus hidden in the trend of the Coulomb term.
Notice, that the observed opposing trends of the two largest depicted terms,
the kinetic and nuclear attraction energies,
are in agreement with the virial theorem.
Since this requires the
sum of the potential energy terms ($E_\text{nuc} + E_\text{e2e}$)
to be related to the kinetic energy by a factor of $-2$,
the same should hold true for the slopes of these as $k$ is varied.
Neglecting the electron-electron interaction
this relationship can indeed be roughly observed for the two other curves.

\begin{figure}
	\centering
	\includegraphics[scale=.8]{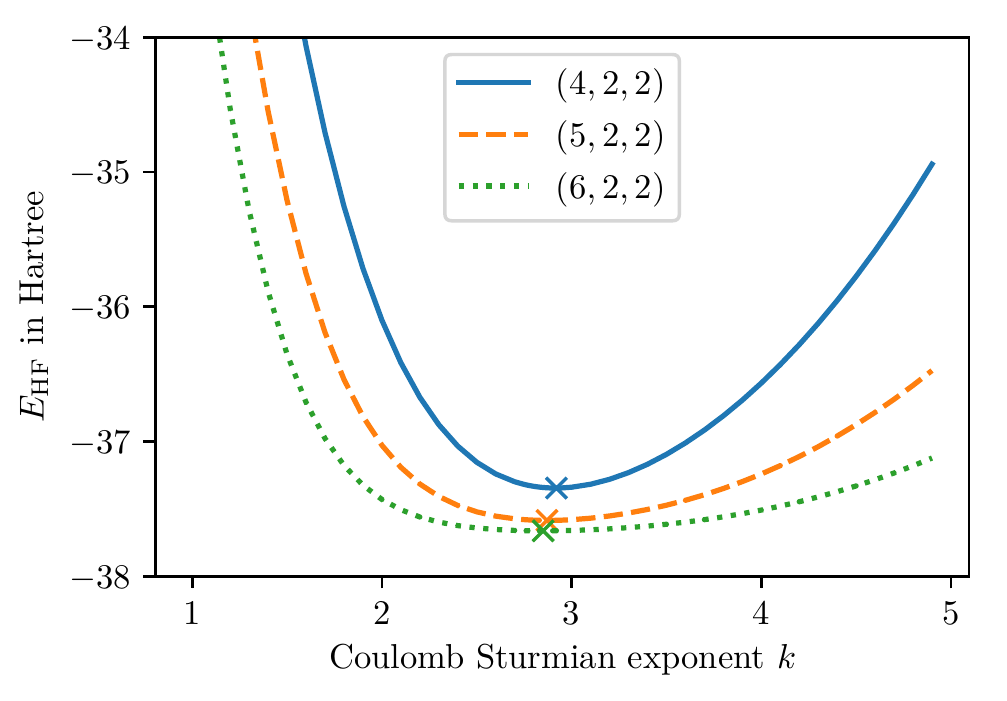}
	\caption
	{
		Plot of the \UHF energies of carbon
		versus the Coulomb Sturmian exponent $\kexp$
		in the $(4,2,2)$, $(5,2,2)$ and $(6,2,2)$ basis sets.
		The minimum-energy exponent $\kopt$
		for each basis set is marked by a cross.
	}
	\label{fig:EHF_vs_k}
\end{figure}
Summing up all energy contributions leads to curves such as
Fig.~\ref{fig:EHF_vs_k},
which shows the total Hartree-Fock energy versus
the Coulomb Sturmian exponent $\kexp$.
From our discussion of the individual terms
it is apparent that at small values for $\kexp$
the increase in nuclear attraction energy dominates,
such that the \HF energy increases rapidly.
At large distances the kinetic energy and electron-electron
interaction terms win giving rise to convex curves
for the plot $E_\text{HF}$ versus $\kexp$.
The shape of these curves depends
on the maximal principle quantum number $\nmax$ of the basis set.
Whilst a $(4,2,2)$ \CS basis gives rise to the deepest minimum,
for $(5,2,2)$ and $(6,2,2)$ the energy versus exponent curves
become visibly flatter close to the optimal exponent (around $\kexp = 2.8$).
Since $\kexp$ only occurs in the radial part of the \CS basis functions
and larger values of $\nmax$ imply
that the set of all radial functions $R_{nl}$ becomes more complete,
the value of $\kexp$ gets less important.
Notice that not all parts of the energy versus exponent curves
are equally dependent on $\nmax$.
As a result the optimal exponent $\kopt$
depends on $\nmax$ as well
and larger basis sets give rise to smaller values for $\kopt$.
This can be rationalised by taking the plots of the energy terms
in Fig.~\ref{fig:EnergyTermsVsK} into account.
The nuclear attraction energy
is influenced by $\kexp$ most strongly and, additionally,
it is (by magnitude) the largest contribution to the \HF energy.
In order to yield the minimal ground-state energy
in a small basis the dominating effect is therefore to
minimise the nuclear attraction energy as much as possible.
As a result the optimal exponent $\kopt$ takes comparatively large values.
As the basis becomes larger a balanced description
of the complete physics becomes possible,
such that the electron repulsion and kinetic effects are described better as well
and thus smaller values for $\kopt$ results.
Moreover the difference in magnitude of the energy terms
rationalises why choosing a \CS exponent larger than $\kopt$
will generally have a lesser influence
on the obtained energy
compared to choosing a too small exponent,
which can be observed in
Fig.~\ref{fig:ErrorHF_vs_shell_k} as well.

Due to the flat structure of the energy versus exponent curves
close to the optimal exponent $\kopt$,
it is not required at high accuracy.
If a highly accurate treatment of a particular system is required,
then increasing $\nmax$ has both a much larger effect
and is computationally cheaper than finding the optimal exponent more accurately.
A good estimate to $\kopt$ for a basis can be usually found by
minimising the \HF energy with respect to $\kexp$
in a smaller basis and then use the obtained value for larger bases as well.
Since such energy curves are convex
and only scalar functions of a single parameter,
this minimisation can be performed effectively
by a gradient-free optimisation algorithm
based on Brent's method~\cite{Brent1972}.
Starting from a reasonable guess for $\kopt$ convergence to the minimum
is usually achieved in around $10$ iterations,
which requires a similar number of
energy computations using the chosen quantum-chemical method
and the chosen \CS basis.
An appropriate procedure for obtaining $\kopt$
for \HF is described in Ref. \onlinecite{Herbst2018Phd}
and has been implemented in \molsturm~\cite{molsturmDesign}.
A selection of optimal exponents for the atoms of the second and
third period can be found in Tables~SI-1 and~SI-2
of the supporting information. 

\subsection{Comparison of optimal Coulomb Sturmian exponents
and the Slater exponents from \citet{Clementi1963}}
\label{sec:ValuesKopt}

\newcommand{\zCl}{\ensuremath \zeta_\text{Clementi}}
Comparing the radial part of a Slater-type orbital~\cite{Slater1930,Helgaker2013}
\begin{equation}
	R^\text{STO}_n(r) = \frac{(2\zeta)^{3/2}}{\sqrt{(2n)!}} (2\zeta r)^{n-1} \exp\left(-\zeta r\right)
	\label{eqn:RadialSlater}
\end{equation}
with the radial part of a \CS function \eqref{eqn:CSdef},
one notices
that the functional form is very similar,
with the Slater exponent $\zeta$ and the \CS exponent $\kexp$
occurring in related terms.
Additionally, the procedure followed by
\citet{Clementi1963} to obtain the Slater exponents $\zCl$
is a variational minimisation of the \HF energy with respect to $\zeta$,
so the same approach we used to obtain $\kopt$.
However, the notable difference between \CS discretisations and \STO
bases is that all \CS functions in a basis share the same $\kexp$,
whereas each function of an \STO basis may employ a different $\zeta$.

\begin{figure}
	\centering
	\includegraphics[scale=.8]{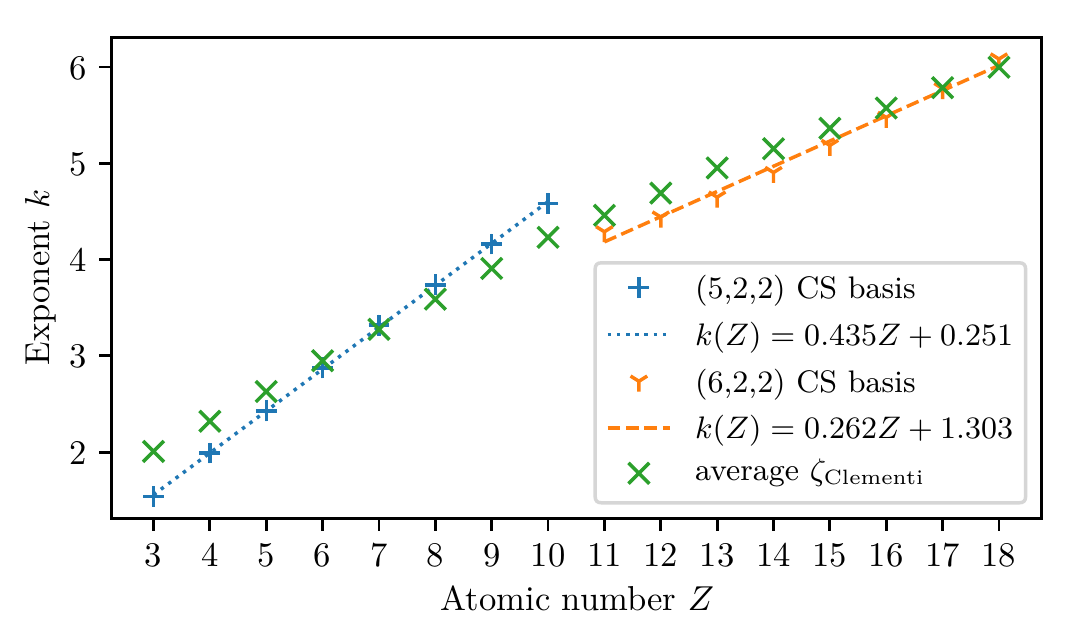}
	\caption{
		Plot of the atomic number versus the optimal Coulomb Sturmian exponent
		$\kopt$ for the atoms of the second and the third period.
		For comparison the occupation-averaged value of the \citet{Clementi1963} optimal
		Slater exponent $\zeta_\text{Clementi}$ are shown as well.
	}
	\label{fig:KoptVsAtnum}
\end{figure}
Instead of directly relating the values of $\kexp$ and $\zCl$,
we therefore plot $\kexp$
versus the occupation-weighted averaged of the $\zCl$ values
corresponding to the occupied orbitals of a respective atom,
see Fig.~\ref{fig:KoptVsAtnum}.
Over the full depicted range the magnitude of $\kopt$ and
the average $\zCl$ stays similar.
Furthermore except the sharp drop going from atom number 10 to 11
the trend of $\zCl$ is more or less reproduced by $\kopt$.
Notice, however, that the trend in both cases is not linear
as can be observed by comparing the data points
to the fitted lines.

With respect to the drop between $Z=10$ and $11$,
two possible causes are plausible.
Firstly, our calculations in the third period employs
a larger \CS basis set compared to the second period.
This was done to provide extra basis functions for the description
of the more electron-rich atoms.
Recalling our discussion in section \ref{sec:ConvK}
related to Fig. \ref{fig:EHF_vs_k},
larger basis sets tend to lead to smaller values for $\kopt$.
The observed drop in Fig.~\ref{fig:KoptVsAtnum} is, however,
much larger than any lowering caused by increasing the basis%
~(see Tables~SI-1 and~SI-2 of the supporting information),
such that additional effects need to be present.
A second aspect to consider is the reduction
of information, which is caused by taking the average of all $\zCl$.
For example when changes in the physics of the electronic structure of the atom
trigger relative adjustments of the exponents $\zCl$,
this is not captured by the average $\zCl$.
Especially when going to a new shell,
\ie when adding a new, more expanded orbital
with only a single electron in it,
the structure of the electron density undergoes
large changes compared to the previous atom:
The inner core electrons become more contracted
while the atomic radius and thus the valence shell expands.
The \STO basis has more degrees of freedom
in form of multiple exponents to adapt to this,
the \CS basis has only one exponent to balance the effects.
This potentially overemphasises some trends,
\eg the compression of the core,
compared to others,
leading to deviations from the trend of the average $\zCl$.

Nevertheless the similarities in the trend between
$\kopt$ and the average $\zCl$,
allows for a physical interpretation of $\kopt$.
Returning to the ideas of \citet{Slater1930},
which were later picked up by \citet{Clementi1963},
one can use the exponents $\zCl$
to define, for each orbital, a shielding parameter $\sigma$,
which indicates how much of the nuclear charge is screened
away by all electrons closer to the core.
The appropriate relationship is
\begin{equation}
	\zCl = \frac{Z - \sigma}{n^\ast},
	\label{eqn:Shielding}
\end{equation}
where $Z$ is the nuclear charge and $n^\ast$
is a function of the principle quantum number, see Ref.~\onlinecite{Slater1930}.
$Z - \sigma$ is sometimes called the effective nuclear charge as well,
giving it the interpretation as
a measure for the remaining charge felt by an orbital.
If we take $\kopt$ to be related to the average $\zCl$,
we can think of $\kopt$ as a measure for the average effective nuclear charge,
which is felt by all electrons.

\subsection{Selecting Coulomb Sturmian basis sets for Hartree-Fock calculations}
As discussed
selecting \CS basis sets for \HF calculations boils down to
selecting a reasonable exponent $\kexp$
together with values for $\nmax$ and $\lmax$
such that the basis does not get too large
and the error in the discretisation of the angular part
as well as the radial part are balanced.

For cases where the $\RMSOl$ plots show a distinct drop
at a particular angular moment, a suitable $\lmax$, which fully captures
the angular part, can be read off.
What remains is to increase $\nmax$ until the radial
part is sufficiently converged as well.
For the examples considered in this work,
$\nmax = 10$ was sufficient to reach a target accuracy of
more than 4 digits in the \HF energy,
which equals a relative error of below $10^{-4}$.
For \ce{Li} and \ce{Be}, where $\lmax = 0$ is sufficient,
this translates to a $(10,0,0)$ basis consisting of only $10$ \CS basis functions.
For the other atoms with a rapid-drop-type $\RMSOl$ plot
a $(10,1,1)$ basis would be required, which has $37$ basis functions.
In our investigation we obtained
rapid-drop-type $\RMSOl$ plots for \HF calculations on closed-shell atoms,
on open-shell atoms with an $S$ ground state term
and for \UHF calculations of any other open-shell atom
if fractional occupation numbers were used.
Given the arguments we outlined in section \ref{sec:HFfractional},
we expect these observation to extend to the other periods.

In case of a decreasing staircase pattern in the $\RMSOl$ plots
one needs to find a balance:
Restricting the \CS basis set using
smaller values of $\lmax$ implies a larger
error in the angular discretisation,
but on the other hand gives rise to more manageable scaling
of the bases set size, see Eq.~\eqref{eqn:NbasAmLimitedBasis}.
This in turn implies that larger values for $\nmax$ can be used
and thus that a more accurate radial discretisation may be obtained.
For example for oxygen in a \UHF calculation with integer
occupation numbers at least $\lmax = 3$ and $\nmax = 10$
is required to reach 5 digits of accuracy compared to the \CBS reference.
This is a basis with the enormous number of $126$ basis functions.
For carbon on the other hand $\lmax = 2$ is sufficient,
such that a $(10,2,2)$ basis with $77$ functions may be used.
On the other hand our discussion linked the occurrence of the decreasing
staircase pattern in the $\RMSOl$ plots to a
breaking of spherical symmetry in the \UHF calculations.
From this point of view
one could argue to still use $\lmax=1$, however.
This will effectively prevent the symmetry breaking
by not providing any higher angular momentum in the basis.
For \UHF calculations of the second and third period
we therefore suggest to stick to $\lmax = 1$.
This approach, however, is not applicable to the higher periods,
since $d$-functions are occupied as well.
For performing accurate \UHF calculations
on open-shell systems in period $4$ and onwards,
either larger $\lmax$ values or techniques to prevent
the breaking of spherical symmetry need to be used.
Alternatively one may still choose $\lmax=2$ and live with the
uneven description of the spherical symmetry breaking in $s$, $p$ and $d$
functions.

Compared to the influence of $\nmax$ and $\lmax$,
the value of the Coulomb Sturmian exponent $\kexp$
only plays a secondary role,
since it does not alter the convergence trends.
Typically it is therefore sufficient to use a value
which is reasonably close to the minimal-energy exponent $\kopt$.
This can for example be achieved by reusing an optimal exponent
from a smaller basis set,
like the exponents provided in the supporting information%
~(Tables~SI-1 and~SI-2).

\section{Outlook}
\label{sec:Outlook}

For judging the convergence properties
of Coulomb Sturmians with respect to quantum-chemical simulations,
Hartree-Fock is without any doubt only the first step.
Nevertheless already at the \HF level this work only represents the first step.
For example restricted open-shell \HF has not been considered at all
so far and similarly we just stated empirical observations.
A more mathematically motivated approach could allow
to deduce rigorous error bounds and potentially allow to understand
whether the observed sublinear convergence for nitrogen
and for the cases with a clear drop in $\RMSOl$
is a general feature,
which would also be encountered for Be and Ne at large enough bases.
Similarly a more quantitative understanding on the deviation
of the convergence rate with respect to choosing a \CS exponent
would be desirable.

With respect to capturing correlation effects,
preliminary work~\cite{Herbst2018Phd} suggests
that the leading order effects can be captured by increasing
$\lmax$ by $1$ --- in agreement with the typical
constructions followed for \cGTO basis sets~\cite{Kong2012,Hill2013,Jensen2013}.
Our plan is to confirm this with a more detailed
discussion in a subsequent publication.
Given that a larger $\lmax$ bound will additionally
increase the basis size beyond the \HF level,
primitive \CS basis functions will probably not be sufficient any more.
With respect to contracted \CS basis sets, however,
the challenge is to design construction schemes,
which do not break the advantageous equivalence of the \CS
basis functions with the hyperspherical harmonics,
which is required for an efficient
evaluation of the \CS ERI integrals~\cite{Avery2013,Avery2015}.
For this reason one should restrict the formation of contracted
\CS functions in a way that all primitives
still share the same \CS exponent $\kexp$.
The availability of contracted \CS basis sets, constructed in such a way,
would furthermore allow to use them in molecular calculations,
which have now gotten into reach due to the recent advances
in evaluating 4-centre electron-repulsion integrals using \CS functions%
~\cite{Avery2012,Avery2013,Avery2015,Avery2017}.

\section{Conclusions}
\label{sec:Conclusion}

In recent years exponential-type basis functions have shown
to be promising alternatives for quantum-chemical simulations~\cite{Guell2008,Hoggan2009}.
From this class of functions Coulomb Sturmians~(\CS) are
particularly appealing.
These functions form a complete one-particle basis and
furthermore their multi-centre electron-repulsion integrals can be evaluated
efficiently~\cite{Avery2012,Avery2013,Avery2015,Avery2017}.
As a result molecular problems could be treated with this basis in the future.
Following along this prospect this work provided a first look
at the construction of \CS basis sets for quantum-chemical calculations.
For this objective a simple and physically motivated construction scheme
for \CS basis sets was suggested and its convergence properties
with respect to atomic calculations at Hartree-Fock~(\HF) level investigated.
A brief outlook towards correlated and molecular calculations was provided as well.

In our construction
a \CS basis set is formed by restricting
the set of possible quantum number triples $(n,l,m)$
using upper bounds $\nmax$, $\lmax$ and $\mmax$ on the
principle, angular momentum and magnetic quantum numbers, respectively.
While the bound on $\nmax$ is required to achieve a finite basis size,
the bounds $\lmax$ and $\mmax$ are optional,
in which case only the usual restrictions between $n$, $l$ and $m$ apply.
The \CS exponent $\kexp$ is common to all \CS functions of the basis and
is fixed as a fourth parameter of a basis.

With respect to the convergence properties
$\kexp$ only effects the convergence rate,
but not the observed convergence trends.
Furthermore, for each choice of $\nmax$ and $\lmax$
an optimal, minimum-energy \CS exponent $\kopt$ can be found.
Deviations from this value, however,
become more and more unimportant as $\nmax$ gets larger.
We have computed some optimal exponent values
for the second and the third period of the periodic table,
which can be found in the supporting information.
From a plot of these $\kopt$ exponents versus the atomic number,
we identified similar trends to
a plot of the average Slater exponents obtained by \citet{Clementi1963}.
Based on these results we suggested a physical
interpretation of the optimal exponents $\kopt$
as a measure for the average effective nuclear charge,
which is felt by the electrons of an atom.

The basis  parameters $\nmax$ and $\lmax$ were identified to independently
influence the convergence of the \CS discretisation
in the radial and angular coordinates, respectively.
Additionally these have a direct influence on the size of the \CS basis,
where introducing a restriction by $\lmax \ll \nmax$
reduces the scaling of the basis set size from cubic in $\nmax$
to linear in $\nmax$.
A key aspect for constructing \CS basis sets is therefore to fix $\lmax$ to a value
causing both a sufficiently good angular discretisation
as well as the smallest basis sizes possible.

For this purpose we introduced the
root mean square occupied coefficient value per angular momentum $l$~($\RMSOl$).
This quantity allows to measure the importance of
a particular angular momentum quantum number $l$
for describing a Hartree-Fock~(\HF) wave function.
Considering the trend of $\RMSOl$ as $l$ is increased
thus either allows to directly select a value for $\lmax$
or help uncover unphysical effects such as the breaking
of spherical symmetry in the \UHF calculations on oxygen and carbon
if integer occupation numbers are used.
It should be noted that the construction of $\RMSOl$ is general
and could be applied to other basis functions of the form
radial part times spherical harmonics, for example \cGTO discretisations.
Due to the completeness of the radial part of the Coulomb Sturmians
the observed $\RMSOl$ behaviour for \CS discretisations
has general character, \ie it should be reproduced
by other basis function types as well.

With respect to the bound $\nmax$, our investigations
indicate that $\nmax = 10$ is sufficient to give rise to
$4$ to $5$ digits of accuracy in the \HF energy
for all our investigated cases.
For lithium and beryllium, $\lmax = 0$ showed to be suitable,
whereas $\lmax = 1$ should be chosen for the other atoms
of the second and third period.
This value assumes, however,
that a potential breaking of spherical symmetry
in \UHF should not be modelled or is prevented using \eg fractional occupation numbers.
As indicated by our $\RMSOl$ plots,
larger values for $\lmax$ are required otherwise
--- the precise value depending on the desired target accuracy.

In all cases considered, the convergence behaviour
of the proposed basis set construction could be interpreted based
on physical arguments.
This includes
challenging cases like the symmetry breaking in some \UHF calculations,
where conversely the obtained $\RMSOl$ plots were used
to gain an understanding of the unusual angular momentum requirements
of such wave functions.
For key basis parameters, such as $\lmax$ or the optimal exponent $\kopt$,
physical interpretations were suggested.
This emphasises the ability of \CS basis sets to capture the
physics of \HF wave functions
and furthermore enables
an intuitive approach to the construction of \CS basis sets.
In light of modern approaches for constructing \cGTO basis sets,
a thorough understanding of convergence properties of
atoms at the \HF level is the prerequisite
for constructing contracted \CS basis sets.
This work enables such progress,
bringing Coulomb Sturmian basis sets one step closer
towards applying them
to correlated quantum-chemical methods and molecular systems.


\section*{Acknowledgments}
The authors thank the group of Prof.~Dr.~Brian Vinter
for providing us with access to their computing facilities
and Guido Kanschat for helpful discussions in the process
of preparing the material.
Michael F. Herbst wishes to express
his thanks to all members of the group of Prof.~Dr.~Brian Vinter
at the Niels Bohr Institute for the hospitality
and many fruitful discussions during visits.
He further
gratefully acknowledges funding by the
Heidelberg Graduate School of Mathematical and Computational
Methods for the Sciences (GSC220).


\bibliography{literature}
\end{document}